\documentstyle[twocolumn,aps]{revtex}
\begin{document}

\title{Measuring Scars of Periodic Orbits }
\author{L. Kaplan\thanks{kaplan@physics.harvard.edu}
\\Department of Physics and
Society of Fellows,\\ Harvard University,
Cambridge, MA 02138\\ \vskip 0.1in
E. J. Heller\thanks{heller@physics.harvard.edu}  \\
Department of Physics and Harvard-Smithsonian
Center for\\ Astrophysics, Harvard University,
Cambridge, MA 02138}
\maketitle

\begin{abstract}

The phenomenon of periodic orbit scarring of
eigenstates of classically chaotic systems is
attracting increasing attention.  Scarring is one of
the most important ``corrections'' to  the ideal
random eigenstates suggested  by random matrix
theory.  This paper discusses measures of scars
and  in so doing also tries to clarify the
concepts and effects of eigenfunction scarring. 
We propose a new, universal scar measure which
takes into account an entire periodic orbit and the
linearized dynamics in its vicinity.  This measure
is tuned to pick out those structures which are
induced in quantum eigenstates by unstable periodic orbits
and their manifolds.  It gives  enhanced scarring
strength as measured by eigenstate overlaps and
inverse participation ratios, especially for
longer orbits.  We also discuss  off-resonance
scars which appear naturally
on either side of an unstable periodic orbit.

\end{abstract}

\section{Introduction}

\subsection{Background}

The modern field of quantum chaology often
associates 
classically chaotic motion on the one hand with 
aspects of random matrix theory  (RMT) on the
other~\cite{bohigas}. These aspects include level
repulsion in the quantum spectrum as given by
the appropriate random matrix
ensemble, Gaussian random wavefunctions with
Bessel correlations, etc.  Since Hamiltonian
dynamics cannot be truly random, numerous recent
contributions to the  field address the many sorts of
``corrections'' to the random matrix approximation.  One
of those corrections is the phenomenon of scarring
of quantum eigenstates by {\it isolated} unstable 
periodic orbits of the corresponding classical
system~\cite{hel84}.
 
In the early 1980's 
MacDonald in unpublished work~\cite{thesis} found clear
evidence of non-isolated marginally unstable
periodic orbits in certain stadium eigenstates (which he
named the ``bouncing ball'' states).   He also
tentatively noted the possible influence of an
{\it isolated, unstable}  periodic orbit on a few
of the calculated eigenstates, but gave no further
attention to this effect. In the subsequent first published
account of numerically computed stadium
eigenstates by MacDonald and Kaufman~\cite{mk}
(which to the authors' knowledge
contains the first eigenstates reported for any 
completely chaotic system),
attention  was  focussed not on
periodic orbit effects but on the nodal structure
of the eigenstates. The  conclusion was that the
expected  pattern of nodal lines for  random
eigenstates had been reached. This early work also
noted   broad agreement with  the Bessel function
form of the eigenstate spatial self-correlation
function for chaotic billiards, as
first discussed and predicted by
Berry\cite{berry}.  The paper~\cite{mk} appeared during the
first epoch of quantum  chaos theory, when much
excitement was being generated by noting similarities
between computed eigenstate properties of
classically chaotic Hamiltonian  systems and
RMT~\cite{bohigas}.

Stronger numerical evidence for the
influence of individual periodic orbits on
eigenstates, together with a theoretical
explanation for scarring  in a chaotic system, was
introduced in 1984\cite{hel84}.  Loosely speaking,
a scar is a concentration  of extra and unexpected
(as compared to the RMT prediction) eigenstate
density near an unstable classical periodic 
orbit. This extra concentration has no classical
analogue, which puts scarring into the family of
quantum localization effects.  A semiclassical
theory for the existence and  strength of scars
was given, using time domain arguments and
dynamics linearized around the periodic orbits. 
This theory has seen a number of  extensions and
applications, including Bogomolny's coordinate
space theory of scarring\cite{bogo} and Berry's
Wigner phase space theory\cite{berryscar}. (The
1984 paper\cite{hel84} had been essentially a
Husimi phase space, or Gaussian wavepacket, theory.)  All
these theories were based on the  linearized
dynamics in the vicinity of a periodic  orbit, but
there were important differences.  For example, an
essential ingredient to the strength of scarring,
the Lyapunov stability exponent of the periodic
orbit, enters only in the  wavepacket
approach\cite{hel84}, while the important observation
of ``knots'' of high density at self-conjugate
(focal) points in coordinate space along the orbit
was made by Bogomolny\cite{bogo}.

Scarring has been shown to affect physical systems
of various sorts\cite{scarexp} and even  the
performance of devices such as a tunnel
diode\cite{diode}. Recently one of us showed that the
decay of metastable states can be strongly affected
by scarring, in that highly anomalous lifetime
distributions are possible depending on where 
decay channels are located with respect to the
shortest periodic orbits of the 
system\cite{kapdecay}.

The  following litany of properties of
eigenfunction scarring has led to some confusion and to
several attempts at developing quantitative 
measures of this phenomenon: (1) The manifestations of
scarring can be subtle or obvious. (2) Measures
of scarring can be basis dependent. (3) Scar-like
structures are found to occur even in artificially
constructed purely random
wavefunctions\cite{helocon}. (4) Statistical
fluctuations allowed by RMT might account for some
apparently scarred states. 

Recent work in our
laboratory\cite{nlscar,sscar,kaplan2} has focussed
much attention on quantifying the phenomenon of
scarring, confirming the  role of the instability
exponent of a given periodic orbit, and further
examining the consequences of these findings,
including experimental
issues and the effects of antiscars. Related
work of ours has noted that  scarring is
concordant with bounds on the ergodicity of
eigenstates as  developed by
Schnirelman\cite{schnirl}, Zelditch\cite{zel}, and
Colin de Verdiere\cite{col}. A review of recent
developments in the theory of scarring
was recently written by one of
the authors\cite{kaplan2}. The present work completes an
important part of the picture by explicitly
addressing the basis dependence of measures of
scarring and  arriving at a universal and optimal
basis for measuring scars, while not diminishing
the utility of simpler and more {\it ad hoc}
measures.

\subsection{Measures}

If eigenstates were ideal random matrix states, 
then all probe states would be equivalent, in that
Gaussian random
statistics in one basis guarantees Gaussian random
statistics in every other. In the RMT literature, it is
sometimes noted that for any given member of the
ensemble there will be a  diagonalizing basis;
however, this basis is non-generic and is itself
randomly varying from one member of the ensemble
to another. One way of approaching the 
corrections to RMT in Hamiltonian  systems  with a
classical analogue is to show that there exist special
bases which are nonrandom and which come from
deterministic dynamical evolution. These special
bases bring the Hamiltonian into a manifestly
non-RMT form.
{\it Any} basis which systematically shows
non-RMT  wavefunction statistics for a classically chaotic system
is thus potentially significant. Seen
in this light, the basis dependence of scar
measures should be expected and even exploited.

One such special basis is that of complex
Gaussian wavepackets. Complex Gaussians have
adjustable position and momentum expectation
values, and satisfy minimum uncertainty conditions in
some system of axes in phase space, making them 
excellent  measuring devices for the structure of
eigenstates in phase space. This basis was the one
chosen in \cite{hel84}.  For Gaussians
centered on periodic orbits, asymptotically exact
($\hbar\to 0$) semiclassical dynamics for a fixed
short time places rigorous non-RMT constraints on
the statistics of  eigenstate projections onto
the Gaussian. Some of the eigenstates (precisely which ones
cannot be specified in this short-time theory)
are then required to  have
large projections onto the periodic orbit centered
Gaussian (these  are the scarred states), while
many more are shown to have anomalously {\it
small} projections (the anti-scarred states). We
review why this is so in the  next section. The
inverse participation ratio (IPR) of such
orbit-centered Gaussian packets is anomalous, and
is governed by the classical  Lyapunov stability
exponent $\lambda$ as 
\begin{equation}
{\rm IPR} \sim {1\over
\lambda}
\end{equation}
for small $\lambda$.
(Note that $\hbar$ does not appear in this scaling,
implying survival of the scarring phenomenon  into
the  classical limit.) Some individual eigenstate
projections onto the Gaussian basis were shown to
be enhanced by  at least $1\over \lambda$ over the
RMT expectation\cite{hel84} (again, exactly which ones is not
known and would require much longer time
information). It is sometimes stated that scar
theory is not a theory of individual eigenstates.
While that is true in many respects, especially of
the energy averaged  approaches such as
Bogomolny's\cite{bogo}, the Husimi (Gaussian
packet)  phase space theory of  scars\cite{hel84}
predicts there must exist individual scarred
states, especially for small
instability exponent $\lambda$.

As versatile as the Gaussian basis is, there are
choices to be made and certain optimizations 
possible which further sensitize the probe basis
to the structures which  classical dynamics
imprints onto the eigenstates of classically
chaotic systems. Before discussing this further,
we note a {\it reductio ad absurdum} which places
restrictions on how far the  refinement of
scarring measures can go.  Some years ago,
Tomsovic and Heller~\cite{tomsovichell} were
successful in  constructing a high quality scarred
eigenstate  of the stadium billiard using only
semiclassical methods (the overlap of the
semiclassical  state with the exact eigenstate being 
0.87).  Using such near-eigenstates  as a probe
basis would lead to  extreme non-RMT behavior in
which all but one eigenstate have small projection
onto the test state. Furthermore, including large
parts of the classical  invariant manifolds
leading far from the region of a given periodic
orbit subverts the idea of a scar of a periodic
orbit.

Fortunately there is  a quite natural stopping 
point in the construction of a test basis: we use only
the linearized dynamics (tangent map) near any
given  periodic orbit  in constructing measures of
scarring.  In this way we arrive at test states
that are understandable in terms of simple
invariant manifold structures of  classical phase
space near periodic orbits.  Although the test
states   can be more complicated than a single
Gaussian, these more sophisticated test states are
still determined by short time  linear dynamics
(of order of the time over which the dynamics
is linearizable, which scales as $|\log \hbar|
/\lambda$). (Recall that a given orbit  can be
highly nonlinear, yet possess a linearizable
tangent map in its vicinity.)

In constructing linear scar measures we still have
a number of choices to make: 
\begin{itemize}

\item What is the uncertainty zone  of the test
state in phase space (in the case of a Gaussian,
this is the uncertainty ellipse in phase space). 
Scar measures will change for example depending on
whether the 
ellipse lies along the stable or unstable manifolds
of the orbit or
away from them.

\item For a given period-one periodic orbit, do we
construct a test state with a Gaussian placed at
one point along the orbit or do we construct a tube
(closely spaced Gaussians on the orbit) of  some
sort, and is this tube to be a coherent superposition of
the Gaussians or an  incoherent one?  (In the case
of a discrete-time map, this corresponds to placing a Gaussian 
at each periodic point of a periodic orbit with
length greater than one.)  

\item Should we take coherent linear combinations
of a given Gaussian and its pre- and post-images,
producing a new test state  consisting of several
different Gaussians at each point along the
orbit?  (This idea leads to the ``universal''
measure of scarring.)

\end{itemize}

\subsection{Brief history of scar measures}

There are several threads in the attempt to make
good  measures of  scarring.  The original
approach\cite{hel84} amounted to projection onto
single Gaussians (the Husimi measure); an $O(1/\lambda)$
enhancement  in the infinite-time average return
probability for a Gaussian  placed on an unstable
periodic orbit was noted for small $\lambda$. This
implies that some eigenstates in specified energy ranges
are systematically
enhanced by $O(1/\lambda)$ in the  periodic orbit regions over the
RMT predictions. Much later it was
realized that this local enhancement has a 
dramatic effect on the tails of the $\vert
\psi\vert^2$ distribution\cite{sscar}.

Any theory of scarring implies some measure of the
effect. The first theory of wavefunction scarring
in position space was developed by
Bogomolny~\cite{bogo}. Bogomolny smoothed the
wavefunction intensity over some small energy
range $\Delta E$ using the semiclassical Green's
function;
scars are represented as   smoothed  sums over
effectively finitely many periodic trajectories of
the system. Bogomolny's semiclassical Green's
function approach is very closely related to our
wavepacket dynamics method, as the semiclassical
Green's function can be obtained from the
semiclassical time-domain propagator by a
stationary-phase Fourier transform. One difference
between the approaches is that Bogomolny envisions
summing over a large number of periodic orbits to
get as close as possible to an energy domain
resolution of order of a mean level spacing. As
mentioned above, in some systems it is indeed
possible to use semiclassical methods to compute
individual eigenstates of the system\cite{ltsc}.
In fact for this purpose one needs information
only about  orbits of period up to the mixing time
(which scales logarithmically with $\hbar$) rather
than the Heisenberg time (which scales as a power
law). However, our aim here is to make predictions
about the distribution of scarring strengths based
only on linearized information around {\it one}
periodic orbit; for this purpose most other orbits
which produce additional oscillations in the
density of states may be treated
statistically~\cite{ott}. It is important to note
in this context that if we are measuring
wavefunction intensities on a given short classical
periodic orbit $\cal P$, then in the semiclassical limit
there are no other short orbits  that come close to this
orbit (on a scale of $\hbar$) in phase space. The
only oscillatory contributions which will need to
be taken into account are from orbits  closely
related to orbits {\it homoclinic} to $\cal P$
[homoclinic orbits are those that approach $\cal P$ at
large negative times, perform an excursion away
from $\cal P$ into other regions of phase space, and
then again approach $\cal P$ at large positive times].
In fact, in the $\hbar \to 0$ limit the periodic
orbit sum for a point $x$ on
a given periodic orbit $\cal P$ can be written
equivalently as a contribution from the orbit $\cal P$
itself plus a sum over trajectories homoclinic to
$\cal P$. Although the two points of view (periodic and
homoclinic sum) are
mathematically equivalent, the homoclinic sum
approach makes explicit the special role of the
orbit $\cal P$ near which we are making measurements.
In the homoclinic return formalism, it is also straightforward
to see that the long-time recurrences of a wavepacket launched
at $x$ are correlated and enhanced
in a way that is determined entirely by the
stability matrix of the short
orbit $\cal P$.

A position space basis, though obviously
physically natural in many measurement situations,
is not generally an optimal one for detecting scar
effects. Unless the periodic point $x$ happens
also to be a focusing point of classical
trajectories near the orbit, only a small fraction
of the total scar strength is captured in the
position basis, and the fraction becomes smaller
as $\hbar$ decreases (or as the energy increases).
An easy way to see this is to notice that the
effects of a classical trajectory in quantum
mechanics {\it generically} extend to a region
around the orbit scaling not as a wavelength but
rather as the square root of a wavelength (and
similarly the affected region scales as the square
root of the total number of channels in momentum
space). Thus, unless either the stable or unstable
manifold of the orbit $\cal P$ at periodic point $x$
happens to be oriented along the momentum
direction, the position space basis will not be
optimal, as reflected in the falling off of the
focusing prefactor with energy in the semiclassical
Green's function (and
similarly the momentum basis will not be optimal,
unless one of the two invariant manifolds is
oriented along the position direction). All this
will become more clear in the exposition of the
following section. In any case, one should keep in
mind that a position space basis can always be
considered as a special limiting case of the
Gaussian wavepacket test state, where the position
uncertainty of the wavepacket becomes comparable
to a wavelength, and the momentum uncertainty
becomes large.

A Wigner phase space analysis of the scarring
phenomenon was given by Berry~\cite{berryscar}.
Berry considered the Wigner function,
again smoothed over an energy interval $\Delta E$
near $E$. Being formulated in phase space, the
approach more closely resembles that of
\cite{hel84}.  Working in Wigner phase
space instead of Husimi space also eliminates the
need to choose the (apparently arbitrary)
eccentricity and orientation of the Gaussian
wavepackets. The downside of Wigner
phase space is the absence of a positivity
condition on the Wigner distribution;
thus the value of the spectral
function cannot be considered as corresponding to
an intensity or a probability of being found near
a certain point $x$ (and random matrix theory is
therefore not applicable). The Husimi function,
which is manifestly positive definite, is
identically a phase space smoothing of the Wigner
distribution over a phase space region scaling as
$\hbar$. The ambiguity in choosing the Gaussian
centered on $x$ over which this smoothing is to be
performed is indeed an important issue, to be
considered carefully in the following. We will see
that to obtain the {\it maximal} scarring effect,
the Gaussian must be chosen to be properly
oriented along the stable and unstable directions
at the periodic point. [An arbitrarily large
wavepacket width is allowed along either of these
directions, with a correspondingly small width in
the orthogonal direction. Also, strong, but
non-maximal, scarring will generally be obtained
for any wavepacket with width scaling as
$\sqrt\hbar$ in both the position and momentum
directions.]

A common limitation of the analyses
\cite{hel84,bogo,berryscar} is that they make no
prediction about the properties of the spectral
fluctuations on scales much smaller than
$\hbar/T_D$, where
$T_D \sim T_P/\lambda$ is the decay time of the
unstable orbit with period $T_P$. Therefore it is not
possible to make quantitative predictions about
{\it specific} individual wavefunction
intensities, participation ratios, etc., without
explicitly doing a Gutzwiller sum over {\it all}
periodic orbits. Even if the sum can be performed,
it is by no means clear that it will converge in
all cases (e.g. in systems where caustics are
important~\cite{caustic}). When the sum does
converge it may produce individual semiclassical
wavefunctions very different from the quantum
eigenstates, due to diffraction and other ``hard
quantum" effects. Furthermore,
such Heisenberg-time
calculations are extremely sensitive to small
perturbations on the system. What one would like
is to be able to say precisely how often a given
single-wavefunction scar strength will appear on a
given orbit, at what energy, and at what parameter
values. In the semiclassical limit, this can in
fact be done using only information about
linearized dynamics near the orbit itself, and, in
some cases, about a few strong isolated homoclinic
recurrences which cannot be treated statistically.

Agam and Fishman~\cite{fishman} define the weight
of a scar by integrating the Wigner function over
a narrow tube in phase space, of cross-section
$\hbar$, surrounding the periodic orbit. Li and Hu
integrate over coordinate space tubes\cite{bambi}.
 Alternatively, de Polavieja, Borondo, and
Benito~\cite{borondo} construct a test state
highly localized on a given periodic orbit using
short-time quantum dynamics.  

Klakow and Smilansky~\cite{smilansky} have used a
scattering approach to quantization to study the
wavefunctions of billiard systems. They treat
carefully the wavefunctions on the Poincare
surface of section, and relate their properties to
scarring in configuration space. Ozorio de
Almeida~\cite{dealmeida} uses the Weyl
representation to establish connections between
classical and quantum dynamics, with particular
application to the semiclassical Wigner function
and scars. Tomsovic~\cite{tomsovic} has used
parametric variation as a new method for studying
scar effects; scars are shown to induce
correlations between wavefunction intensities on a
periodic orbit and the level velocities of these
wavefunctions when certain system parameters are
varied. We also mention the work of Arranz,
Borondo, and Benito~\cite{arranz} who have probed
the intermediate region between regular and
strongly chaotic quantum behavior, and have shown
how scarred states first arise from the mixing of
pairs of regular wavefunctions as $\hbar$ is
decreased (but well before one reaches the
semiclassical limit which is the main focus of the
present work). Finally, several
groups~\cite{voros,baker} have studied
the hyperbolic scar structures associated not only
with the periodic orbit itself but with its
invariant manifolds and homoclinic orbits.
 
In the next section we discuss scarring as
measured by individual Gaussian wavepackets,
which was the basis of \cite{hel84}.
A single localized test state may be optimized to
conform to the classical invariant manifolds in the
vicinity of a scar.
In subsequent
sections we go considerably beyond this measure,
refining our templates to better detect scarring.
In Section~\ref{incoh} we address the apparent
arbitrariness in the choice of a point along the orbit at which
to make the measurement, and in the eccentricity of the
test Gaussian, and eliminate these ambiguities by 
building a wavepacket-averaged measure of scarring.
Following this, in Section~\ref{coh}, we use
coherent linear combinations of the localized test
states as a more sensitive measure. In Section~\ref{offres},
off-resonance scars living on either side of an unstable periodic orbit
are shown to follow naturally from our formalism. In
Section~\ref{hiper} extensions to higher-period orbits and
continuous time are discussed, followed by concluding
remarks in Section~\ref{concl}.

\section{Gaussian wavepacket scarring}

\label{gausscar}

\subsection{Semiclassical dynamics of a Gaussian
wavepacket}

We begin with a review of the original (Gaussian
wavepacket) theory of scarring, as discussed in
detail recently in \cite{nlscar}. In the
course of the discussion, the key concepts of the
autocorrelation function, the short-time spectral
envelope, nonlinear recurrences, and the inverse
participation  ratio will be introduced. We will
also see the inherent limitations of measuring
scar strength using single Gaussian test states,
pointing the way to the construction of improved
``scarmometers'' \footnote{The authors thank
Eugene Bogomolny  for coining this term.} in the
following sections.

Consider an arbitrary (unstable) periodic orbit of
a chaotic system. For the purpose of simplifying
the exposition, and without loss of generality, we
take the periodic orbit to be a fixed point of a
discrete-time area-preserving map on a
two-dimensional phase space. If the periodic orbit
in question is in fact a higher-period orbit of
such a map, or is an orbit of a continuous-time
dynamics in two spatial dimensions, we can reduce
the problem to the preceding case by iterating the
original map, or by taking a surface of section
map, respectively.\footnote{The issue of higher-period
orbits and continuous time will be
addressed explicitly in Sections~\ref{incoh} and
\ref{hiper}.}

We start with a fixed point  at the origin of
phase space. Furthermore, we can take the stable
and unstable directions at the fixed point to be
vertical ($p$) and horizontal ($q$), respectively
(we can always get the local dynamics into this
form by first performing a canonical
transformation on the coordinates). Then the only
parameter describing the local (linearized)
dynamics near the orbit is $\lambda$, the
instability exponent for one iteration of the
orbit. Locally, the equations of motion are given
by
\begin{eqnarray}
\label{eqmo} q \to q' &=&e^{\lambda t}q \nonumber
\\ p \to p'&=&e^{-\lambda t}p \,.
\end{eqnarray}

We now turn to the construction of a test state
which can be used to measure the intensity of
eigenstates near the chosen periodic orbit. An
obvious choice is a Gaussian wavepacket centered
on the fixed point:
\begin{equation}
\label{wavepkt} a_\sigma(q)=\left({1 \over \pi
\sigma^2 \hbar}\right)^{1/4}
 e^{-q^2/2\sigma^2\hbar} \,.
\end{equation} This is a minimum-uncertainty state
centered at the origin of phase space, with width
$\sigma\sqrt\hbar$ in the $q-$direction and
$\sqrt\hbar/\sigma$ in the $p-$direction. $\sigma$
is at this stage an arbitrary parameter: $\sigma^2$ is the
aspect ratio of the phase-space Gaussian,
typically chosen to be of order unity. Ambiguity
in the choice of $\sigma$ is an important issue
that we will return to at the beginning of the
next section.

Eigenstate overlaps with our test state will
provide a good measure of eigenstate intensities
near the periodic orbit; however, we find it
useful to begin by working in the time domain (our
results will then be applied to eigenstate
properties in the following subsection).

For small enough $\hbar$, the wavepacket
$|a_\sigma\rangle$ and its short-time iterates are
contained well within the linear regime. As long
as the wavepacket stays in the phase space region
surrounding the periodic orbit in which the
linearized equations of motion Eq.~\ref{eqmo}
apply, the evolution of the wavepacket is
completely semiclassical, given simply by the
stretching of the $q-$width parameter $\sigma$.
More explicitly, at short times we have
\begin{equation}
\label{wpevol} U^t |a_\sigma\rangle \approx U_{\rm
lin}^t |a_\sigma\rangle = e^{-i\phi t} |a_{\sigma
e^{\lambda t}} 
\rangle \,,
\end{equation} where $U$ is the unitary operator
implementing the quantum discrete-time dynamics,
$U_{\rm lin}$ represents the quantization of the
linearized behavior near the periodic orbit, and
$t$ is time, measured in units of a single mapping.
Here
$-\phi$ is a phase associated with one iteration
of the periodic orbit: it is given by the
classical action in units of $\hbar$, plus Maslov
indices if appropriate.

The autocorrelation function of the wavepacket is
defined as the overlap of the evolved wavepacket
with itself:
\begin{equation} A(t)=\langle a_\sigma | U^t |
a_\sigma
\rangle \,,
\end{equation} which at short times is seen from
Eqs.~\ref{wavepkt},~\ref{wpevol} to be
\begin{equation}
\label{shortcorr} A_{\rm lin}(t) = e^{-i\phi t}
\langle a_\sigma | a_{\sigma e^{\lambda t}}
\rangle = {e^{-i\phi t} \over \sqrt{\cosh(\lambda
t)}} \,,
\end{equation} by performing a simple Gaussian
integration. The `lin' subscript indicates that
Eq.~\ref{shortcorr} describes the piece of the
autocorrelation function coming from the
linearized dynamics around the periodic orbit. For
a weakly unstable orbit (small $\lambda$), $A_{\rm
lin}(t)$ is slowly decaying, with strong
recurrences happening for the first $O(1/\lambda)$
iterations of the orbit. We note that the
short-time autocorrelation function $A_{\rm
lin}(t)$ is $\sigma$-independent, a fact that will
prove important later on.

At longer times, namely beyond the log time, which
scales as
\begin{equation}
\label{logtime} T_{\rm log} \sim {\log fN \over
\lambda}\,,
\end{equation} the wavepacket leaves the
linearizable region and nonlinear recurrences
begin to dominate the return probability. Here $N$
is the total number of  Planck-sized cells in the
accessible phase space (also equal to the
dimension of the effective quantum mechanical
Hilbert space), and $f$ is the fraction of this
phase space (typically $O(1)$) in which the
linearized equations of motion (Eq.~\ref{eqmo})
apply. The nonlinear recurrences correspond to a
piece of the wavepacket leaving the linear regime
along the unstable manifold, undergoing
complicated dynamics far from the periodic orbit,
and eventually coming back along the stable
manifold to intersect the original wavepacket.
Semiclassically, these recurrences are given by a
sum over points homoclinic to the original
periodic orbit (i.e. points that approach the
periodic orbit both as
$t \to +\infty$ and  as $t \to -\infty$). 

Because the long-time homoclinic orbits come back
with complicated accumulated phases, and the
number of these recurrences grows exponentially
with time, one might expect the total long-time
return amplitudes to be given by Gaussian random
variables. In fact, however, contributions from
all homoclinic points lying on a single homoclinic
orbit (i.e. those that are exactly time-iterates
of one another) come back in phase with each
other, giving rise to short-time correlations in
$A(t)$ for large $t$ \cite{nlscar}. These
correlations are related to the short-time
dynamics of the original Gaussian wavepacket. In
fact, we can write the return amplitude at times
$T_{\rm log} \ll t \ll T_H$  ($T_H=N$ is the
Heisenberg time, where individual eigenstates
begin to be resolved) as a convolution 
\begin{equation}
\label{convol} A(t)=\sum_\tau A_{\rm rnd}(\tau)
A_{\rm lin}(t-\tau) \,.
\end{equation} Here $A_{\rm lin}$ is the
short-time return amplitude, and $A_{\rm rnd}$ has
the statistical properties of an uncorrelated
random Gaussian variable. In effect, random
recurrences due to Gaussian fluctuations must have
``echoes'' that mirror the initial short time 
decay, since the recurrences re-load the initial
state. Also,
\begin{eqnarray}
\label{rnd} <A_{\rm rnd}(\tau)> & = & 0 \nonumber
\\ <A^\star_{\rm rnd}(\tau)A_{\rm rnd}(\tau')> & =
&
 {1 \over N} \delta_{\tau \tau'}\,.
\end{eqnarray} The prefactor $1/N$ provides the
proper classical normalization: in the absence of
interference effects, the probability to come back
is equal to the probability for visiting any other
state in the Hilbert space. The average in
Eq.~\ref{rnd} is taken over long times $\tau$,
$T_{\rm log} \ll \tau \ll T_H$, and/or over an
ensemble of systems which all have the same
linearized dynamics around our chosen periodic
orbit. In either case, the total size of the
Hilbert space $N$ ($=1/h$ for a phase space area
normalized to unity) has been assumed to  be
large. We then obtain
\begin{eqnarray}
\label{longcorr} <A(t)> & = & 0 \nonumber \\
<A^\star(t)A(t+\Delta)> & = & {1 \over N} \sum_s
A^\star_{\rm lin}(s) A_{\rm lin}(s+\Delta)  \,.
\end{eqnarray} At times beyond the Heisenberg
time, this gets modified \cite{nlscar} to
\begin{equation}
\label{longcorr2} <A^\star(t)A(t+\Delta)> = {F
\over N} \sum_s A^\star_{\rm lin}(s) A_{\rm
lin}(s+\Delta)  \,.
\end{equation}
$F$ is a factor associated with the discreteness
of the eigenstates: it is
$3$ for real eigenstate--test state overlaps and
$2$ for complex overlaps.

The long-time autocorrelation function is thus
self-correlated on a scale
$\Delta \sim \lambda^{-1}$. Qualitatively, this
can be understood on a purely classical level:
once probability happens to come back to the
vicinity of a weakly unstable periodic orbit, it
tends to stay around before leaving again. On the
other hand, the overall enhancement in the total
return probability at long times:
\begin{equation}
\label{enhprob} <|A(t)|^2> = {F \over N}
\sum_{s=-\infty}^\infty {1\over \cosh(\lambda s)} \,, 
\end{equation} obtained by combining the general
expression Eq.~\ref{longcorr2} with the short-time
overlap dynamics of the Gaussian wavepacket
(Eq.~\ref{shortcorr}), is fundamentally an
interference phenomenon, and signals a kind of
quantum localization, as we shall see next.
Note that in the limit $\lambda \to 0$
(weak instability) we have
\begin{equation} 
<|A(t)|^2>\to {\pi F \over \lambda N};
\end{equation} 
i.e. the enhancement factor in the long-time return probability
is proportional to $\lambda^{-1}$\cite{hel84}.

\subsection{Local density of states}

We now define $S(E)$ to be the fourier transform
of the autocorrelation function,
\begin{equation}
\label{spectrum} S(E)={1 \over 2\pi}
\sum_{t=-\infty}^{+\infty} A(t) e^{iEt}\,.
\end{equation} For a non-degenerate spectrum, it
is easy to see (by inserting complete sets of
eigenstates) that
\begin{equation}
\label{linespec} S(E)=\sum_n |\langle n
|a_\sigma\rangle|^2 \delta(E-E_n) \,,
\end{equation} where $E_n$ are the eigenvalues of
the dynamics, and $|n\rangle$ are the
corresponding eigenstates. Thus, we obtain the
local density of states at the wavepacket
$|a_\sigma\rangle$ by fourier transforming its
autocorrelation  function $A(t)$. Cutting off the
sum in Eq.~\ref{spectrum} at $\pm T_{\rm log}$, or
equivalently by including only linearized dynamics
around the periodic orbit, we obtain {\it the
smoothed local density of states}:
\begin{equation}
\label{smspectrum} S_{\rm lin}(E)=\sum_t A_{\rm
lin}(t) e^{iEt}\,,
\end{equation} an envelope centered at quasienergy
$E=\phi$ (see Eq.~\ref{shortcorr}),
 of width $\delta E \sim \lambda$, and of height
$\sim \lambda^{-1}$ (a factor of $2 \pi$ has been
inserted into the definition of $S_{\rm lin}$ for
future convenience). $E=\phi$ is the analogue of
the EBK quantization condition for  integrable
systems; here, because of the instability of the
orbit, scarred states can live in an energy range
of $O(\lambda)$ around the optimal energy. States
with energy more than $O(\lambda|\log \lambda|)$ away from
resonance tend to be {\it antiscarred} (i.e. they
have less than expected intensity at the periodic
orbit).

Now long-time (nonlinear) recurrences as in
Eq.~\ref{convol} lead to fluctuations under the
short-time envelope in the full spectrum $S(E)$.
Because these recurrences involve a random
variable {\it convoluted} with the short time
dynamics, in the energy domain we obtain random
fluctuations {\it multiplying} the short-time
envelope. (It is easy to see physically that the
random oscillations must multiply the smooth
envelope: if they were merely added to it, the
total spectrum would become negative away from the
peak of the envelope.) Finally, at the Heisenberg
time $T_H = N$, individual states are
resolved~\cite{nlscar,sscar}, and we see a line
spectrum with a height distribution given by 
\begin{equation}
\label{ran} I_{n a_\sigma} \equiv |\langle
n|a_\sigma\rangle|^2 = r_{an} S_{\rm lin}(E_n) \,,
\end{equation} where $r_{an}$ are random variables
(with mean $<r_{an}>={1/N}$) drawn from a
chi-squared distribution of one degree of freedom
(two degrees of freedom for complex
$\langle n|a_\sigma\rangle$). Thus, in the end we
obtain a random (Porter-Thomas) line spectrum $S(E)$, all
multiplying the original linear envelope.

Before concluding this review, we mention the
notion of an {\it inverse participation ratio}
(IPR), a very useful measure for studying
deviations from quantum ergodicity. We define
\begin{equation} {\rm IPR}_{a_\sigma} = N \sum_n
I_{n a_\sigma}^2 = N \sum_n |\langle
n|a_\sigma\rangle|^4 \,.
\end{equation} (Note that $\sum_n I_{n
a_\sigma}=1$ by normalization.)  Being the first
non-trivial moment of the eigenstate intensity
($I_{n a_\sigma}$) distribution, the IPR provides
a convenient one-number measure of the strength of
scarring (or any other kind of deviation from
quantum ergodicity). The IPR would be unity for a
wavepacket that had equal overlaps with all the
eigenstates of the system; the maximum value of
$N$ is reached in the opposite (completely
localized regime), when the wavepacket is itself a
single eigenstate. Random matrix theory predicts
an IPR of $F$, the strong quantum ergodicity
factor defined above in Eq.~\ref{longcorr2}.

From Eqs.~\ref{spectrum},~\ref{linespec}
we see that
\begin{equation}
\label{iprsum} {\rm IPR}_{a_\sigma} = \lim_{T \to
\infty} {N \over T} \sum_{t=0}^{T-1} |A(t)|^2 \,;
\end{equation} as one might expect, localization
is associated with an enhanced return  probability
at long times. Now from Eq.~\ref{enhprob} we see
that scar theory predicts an enhancement in the
IPR over random matrix theory:
\begin{eqnarray}
\label{scaripr} {\rm IPR}_{a_\sigma} & = & F
\sum_s {1\over \cosh(\lambda s)} \\ & \to & F {\pi
\over \lambda} \,,
\end{eqnarray} where in the last line the limit of
small $\lambda$ has been taken.
($F$, as before, is $3$ or $2$,
depending on whether the states are real or
complex, respectively.) The IPR thus decomposes
into a product of two contributions: the shape of
the short-time envelope coming from the linear
dynamics around the periodic orbit, and a quantum
fluctuation factor $F$, as predicted by
Porter-Thomas statistics.

\subsection{Limitations}

The analysis of the previous two subsections has
been extensively tested in numerical
studies~\cite{nlscar,sscar}, which show that the
statistical  properties of eigenstate overlaps
with Gaussian wavepackets can indeed be described
by the scar theory. However, there are inherent
limitations in this approach. An obvious one is
the ambiguity in the choice of wavepacket width
$\sigma$. A wavepacket of any width can be used
(as long as it and its short-time iterates are
well-contained in the linearizable region, which
condition is always satisfied for small enough
$\hbar$), resulting in the same short-time
overlaps, and thus in identical smoothed spectra
$S_{\rm lin}$. The IPR is also expected to be
enhanced by the same factor for each such
wavepacket, depending only on the decay exponent
of the periodic orbit itself. It seems intuitively
clear that a better measure of scarring should be
obtainable by appropriately combining information
from wavepackets of all different aspect ratios,
thus looking at a hyperbolic phase-space region
surrounding the stable and unstable manifolds of
the periodic point. Such a test state for
measuring scars would incorporate knowledge of the
full linearized dynamics around the periodic
point, not just knowledge about the location of
the periodic point itself.

The ambiguity and apparent arbitrariness of the
preceding definition of scarring seems even more
pronounced in the case of a higher-period orbit of
a map, or for a periodic orbit of a
continuous-time dynamics. In either of those
cases, the analysis above can be performed at {\it
any} periodic point lying on the orbit. Yet it is
known from experience that scars tend to live not
at one periodic point only but along the entire
orbit. Here, also, more information could 
presumably be gained by looking at the behavior of
an eigenstate near {\it all} points on a  periodic
orbit instead of one only, thus obtaining a fuller
measure of wavefunction scarring.

The preceding ambiguities will, in the following
sections, point us towards a {\it universal}
measure of wavefunction scarring (in the regime of
small $\lambda$, where the effect is expected to
be significant), a measure which takes full
advantage of the {\it entire} periodic orbit and
the {\it full linearized dynamics} in the vicinity
of this orbit. In the process, we will see how the
insights of various earlier contributors to this 
field\cite{fishman,saraceno,borondo,tomsovic,voros}
can be incorporated into the resulting general
approach.

\section{Wavepacket intensity averaging}
\label{incoh}

\subsection{Density matrix test states}

Consider again the fixed point of a classical
area-preserving map, as introduced in
Eq.~\ref{eqmo}. Given the apparent arbitrariness
in the choice of wavepacket which came out of our
discussion in the previous section, it seems
natural to extend our measure of scarring,
replacing the pure Gaussian test state with a
density matrix which gives weight to Gaussians of
all widths:
\begin{equation}
\label{rhodef}
\rho = {\cal N} \int dt \, e^{-t^2/T_0^2} \,
|a_{\sigma e^{\lambda t}} \rangle \langle
a_{\sigma e^{\lambda t}} | \,.
\end{equation} We choose the exponent $\lambda$ to
be the instability exponent of the  unstable orbit
(Eq.~\ref{eqmo}), but this choice is in fact
arbitrary, and $\lambda$ can be reabsorbed into
the overall normalization factor $\cal N$ and the
time cutoff $T_0$. In the absence of the cutoff
$T_0$, the hyperbolic test state would be
completely scale invariant, giving equal weight 
to Gaussians of all aspect ratios, from tall and
thin to short and wide. The cutoff is, however,
necessary because the linearized dynamics of
Eq.~\ref{eqmo} is in fact only valid in a finite
classical region around the periodic orbit (and it
also eliminates possible normalization
difficulties).  We note that Eq.~\ref{rhodef} 
is an {\it incoherent } superposition of 
wavepackets of different width, designed to 
give an unbiased measure of a scar (where the
arbitrariness of a given choice of $\sigma$
is removed).

Let the classical region in which the dynamics is
linearizable be given by a square in phase space,
with area $A$ (the exact shape and area are not
important, as we will soon see). Then choosing
$\sigma=1$, we see that the evolved width $\sigma
\sqrt\hbar
 e^{\lambda t}$ reaches the edge of the
linearizable region at time 
\begin{equation} T_0= {1 \over 2 \lambda} \log {A
\over 4 \hbar} \,.
\end{equation} (Up to constants, $T_0$ is the same
as the log time  discussed above in
Eq.~\ref{logtime}.) We see that a factor of order
one  ambiguity in the definition of the region $A$
will lead  only to an additive ambiguity in the
cutoff time $T_0$, irrelevant in the semiclassical
limit $A \gg \hbar$. Of course the condition that
the Gaussian just touching the boundary should be
suppressed by a factor of $1/e$ (as opposed to
$1/e^2$, $1/e^3$, etc.) is still  somewhat
arbitrary, leading us to the more general form
\begin{equation}
\label{t0def} T_0= {c \over 2 \lambda} \log {A
\over 4 \hbar} \,,
\end{equation} where $c$ is now an arbitrary
constant of order unity. In any case, the
ambiguity we previously had in the choice of
$\sigma$ (which could be anywhere from $\sqrt
\hbar$ to $1/\sqrt \hbar$, a huge range in the
semiclassical regime $\hbar \ll 1$) has now been 
reduced to a factor of order one constant $c$ in
the definition of $\rho$.

In the following section, where  we consider
{\it coherent} superpositions of Gaussian
wavepackets, we may wish to use a more stringent
criterion for the cutoff time $T_0$, taking into
account the form of the leading nonlinearity of
the dynamics near the periodic orbit. Thus,
consider the worst-case scenario, where the stable
and unstable manifolds both curve quadratically as
we move away from the periodic orbit. Then in
order for the curvature to be unimportant as the
Gaussian stretches along one of the two manifolds,
we may require that the distance by which the
unstable manifold deviates from the horizontal
line at position $q$ be less than the vertical
width (momentum uncertainty) of a state with
position width $q$. This means $ O(q^2) < \delta p
\sim
\hbar /q$, so the maximum distance $q$ for which
this holds scales as $\hbar^{1/3}$. Assuming the
same situation obtains along the stable manifold,
we obtain that the linearizable area scales as
$\hbar^{2/3}$, i.e. $A=A_0^{1/3}\hbar^{2/3}$ for
some classically selected area $A_0$. Then $\log
(A/\hbar)={1 \over 3} \log (A_0/\hbar)$, i.e. we
may take the size of the linearizable region to be
the $\hbar-$independent value
$A_0$, provided we also replace $c \to c/3$.
Because of the $O(1)$ ambiguities already present
in the choice of $c$, we will not dwell here on
the numerical values appropriate to various
systems. In any case, as we will discuss when
subjecting our results to numerical tests in the
following section, there is always a tradeoff
between larger $c$ leading to more localization
and smaller $c$ giving more precise agreement with
analytical predictions.

Coming back now to Eq.~\ref{rhodef}, we notice
that we could instead have chosen a hard cutoff
for the sum over Gaussians, e.g.
\begin{equation}
\rho' = {\cal N'} \int_{-T_0}^{T_0} \, dt \,
|a_{\sigma e^{\lambda t}} \rangle \langle
a_{\sigma e^{\lambda t}} | \,.
\end{equation} This would not qualitatively affect
our discussion either here or in the following
sections (particularly Section~\ref{coh},
 where we discuss coherent superpositions of
Gaussian  test states). The form of
Eq.~\ref{rhodef}, and its extensions which will
follow in future sections, is however convenient
because it allows for relatively straightforward
analytical calculations.

Another important property of Eq.~\ref{rhodef} is
that $\rho$ has the same form in momentum space as
in configuration space, as can be seen easily by
noting that the momentum width
$\sigma_p=\sigma_q^{-1}=\sigma^{-1}e^{-\lambda t}$
and that Eq.~\ref{rhodef} is manifestly invariant
under
$t \to -t$. Thus, the stable and unstable
manifolds of the hyperbolic point are treated
symmetrically in our definition.

The idea of averaging overlap intensities can of
course be extended to resolve another difficulty
we encountered at the end of the previous section,
namely the apparent ambiguity in treating periodic
orbits that are not fixed points. For a period
$T_P$ orbit of a map, we write
\begin{equation}
\label{orbsum}
\rho={1 \over T_P}
\sum_{p=0}^{T_P-1} \, |a_{x_p,\sigma} \rangle
\langle a_{x_p,\sigma}| \,,
\end{equation} where $|a_{x_p,\sigma} \rangle$ is
a wavepacket of width $\sigma$ along the unstable
manifold, but centered at periodic point $x_p$
instead of at the origin. Similarly, for a
continuous-time dynamics, we can write
\begin{equation}
\label{tube}
\rho= {\cal N} \int dx  \, |a_{x,\sigma,\sigma_x}
\rangle
\langle a_{x,\sigma,\sigma_x}| \,,
\end{equation} where the $x$ coordinate
parametrizes the periodic orbit in phase space,
and at each periodic point the wavepacket is
chosen to have width $\sigma_x$ along the
direction of the orbit and width $\sigma$ in the
unstable direction at that point on the orbit:
\begin{eqnarray} a_{x,\sigma,\sigma_x}(x',y') \sim 
\exp[&-&(x'-x)^2/\sigma_x^2 \hbar-y'^2/\sigma^2
\hbar \nonumber \\
 & + & i p_x (x'-x)/\hbar] \,.
\end{eqnarray} Here $(x,p_x)$ are the position of
a phase space point on the periodic orbit and the
corresponding momentum, while $y'$ is a coordinate
along the unstable manifold of the orbit at point
$(x,p_x)$. Eq.~\ref{tube} has been written down
already in Ref.~\cite{nlscar}, and a connection
was made there to the phase-space tubes of Agam,
Fishman et al.~\cite{fishman}.

Now the orbit averaging of
Eqs.~\ref{orbsum},~\ref{tube} can of course be
combined with the width averaging introduced in
Eq.~\ref{rhodef}: thus, in the case of a map we
may write
\begin{equation}
\label{combined}
\rho={\cal N} 
\sum_{p=0}^{T_P-1} 
\int dt \, e^{-t^2/T_0^2}
\, |a_{x_p,\sigma e^{\lambda t}} \rangle \langle
a_{x_p,\sigma e^{\lambda t}}| \,.
\end{equation}

Here one may legitimately ask why we perform
separately the averaging at each periodic point:
we could instead have made use of the orbit
dynamics and obtained Gaussian wavepackets
centered at each of the periodic points starting
with one wavepacket only and allowing it to evolve
according to the linearized laws of motion. More
explicitly, if we take 
$|a\rangle$ to be a Gaussian wavepacket of width
$\sigma$ along the unstable direction, centered at
periodic point $x_0$, we may construct a dynamical
density matrix
\begin{equation}
\label{rhodyn}
\rho_{\rm dyn} = \sum_{t=-\infty}^{+\infty} \,
e^{-t^2/T_P^2 T_0^2} \, |a_{\rm lin}(t)\rangle
\langle a_{\rm lin}(t)| \,.
\end{equation} Here $|a_{\rm lin}(t)\rangle$ is
the original Gaussian evolved in accordance with
the {\it linearized} dynamics: for example, if $t$
is an integer multiple of the period $T_P$, then
$|a_{\rm lin}(t)\rangle$ is centered at the same
periodic point as 
$|a\rangle$, but with width $\sigma e^{\lambda
t/T_P}$.
$T_0$ is defined as before (Eq.~\ref{t0def}),
using the full instability exponent $\lambda$ for
one iteration of the {\it entire} primitive orbit.
$\lambda/P$ is the exponent {\it per time step};
hence the factor of $T_P^2$ in Eq.~\ref{rhodyn}.

For small $\lambda$, where not much stretching has
taken place over one period of the orbit, not much
difference exists between the averaging methods of
Eq.~\ref{orbsum} and Eq.~\ref{rhodyn}. We return
to this connection between (linearized) dynamics
and improved test states in Section~\ref{coh},
where coherent superpositions of Gaussian
wavepackets are discussed. Here it suffices to
note that because all wavepackets being averaged
over in $\rho_{\rm dyn}$ are exact time-evolutes
of one another (at least in the linear
approximation), they all have exactly the same
local density of states and inverse participation
ratio. In fact this absence of real averaging is
there even for the full matrix $\rho$ of
Eqs.~\ref{orbsum},~\ref{tube},~\ref{combined}, in
the limit of small
$\lambda$, as we shall see next.

\subsection{Measures of scarring from incoherent
averaging}

Our measure for the strength of scarring for a
given eigenstate $|n
\rangle$ is now simply
\begin{equation} I_{n \rho} \equiv \langle n |
\rho | n \rangle \,.
\end{equation} We can construct a
wavepacket-averaged local density of states
analogous to Eq.~\ref{linespec}
\begin{equation} S_\rho(E)=\sum_n I_{n \rho}
\delta(E-E_n) \,,
\end{equation} and a corresponding inverse
participation ratio
\begin{equation} {\rm IPR}_\rho = N \sum_n I_{n
\rho}^2 \,.
\end{equation} Notice that $S_\rho(E)$ is nothing
other than a weighted sum of the densities $S(E)$
of Eq.~\ref{linespec}, and thus follows the same
linear envelope $S_{\rm lin}$ which we have
discussed in the previous section. The only thing
possibly different about $S_\rho(E)$ are the
oscillations under this envelope. To understand
how these oscillations in the averaged local
density of states ${\rm IPR}_\rho$ might differ
from the Porter-Thomas fluctuations one finds for
a single wavepacket, we need to study correlations
between local densities of states for different
wavepackets centered on the same periodic orbit.

In general, given two wavepackets $|a\rangle$ and
$|b\rangle$, we can define a long-time averaged
transport probability $P_{ab}$ \cite{wqe} as
\begin{equation} P_{ab} = \lim_{T \to \infty} {1
\over T} \sum_{t=0}^{T-1} |\langle a | U^t |b
\rangle |^2 \,.
\end{equation} For a nondegenerate spectrum we
easily see
\begin{equation} P_{ab} = \sum_n \, |\langle a |
n\rangle|^2 \, |\langle b | n\rangle|^2 = \sum_n
I_{na} I_{nb} \,.
\end{equation}  In particular, the IPR as defined
in the preceding section corresponds to the
special case $|a\rangle=|b\rangle$:
\begin{equation} {\rm IPR}_a = N P_{aa}  = N
\sum_n I_{na}^2 \,.
\end{equation} The $P_{ab}$ can be thought of as
the covariance matrix of the densities of states
for different wavepackets, with $P_{aa}$ being the
variances or diagonal matrix elements; the
correlation between two densities of states is then
given by
\begin{equation}
\label{cabdef} C_{ab} = {P_{ab} \over \sqrt{
P_{aa} P_{bb} }} \,.
\end{equation}

We begin with the simplest case, where wavepackets
$|a\rangle$ and $|b\rangle$ are exact time
iterates of one another: $|b\rangle=|a(t)\rangle$
for some time $t$. There, of course
$P_{ab}=P_{aa}=P_{bb}$, and the correlation is
unity (the two local densities of states $S(E)$
are identical). More explicitly (from
Eq.~\ref{scaripr}) we have in this case
\begin{equation} P_{ab}=P_{aa}=P_{bb}= {F\over N}
g(\lambda) \,,
\end{equation} where $F/N$ is the RMT prediction
for the quantum long-time return probability and 
\begin{equation} g(\lambda) =\sum_s |\langle a
|U^s_{\rm lin}|a \rangle|^2=
\sum_s {1\over \cosh(\lambda s)}
\end{equation} is a scarring IPR enhancement
factor (see Eq.~\ref{scaripr}). In the last
equality the periodic orbit in question has been
taken to be period one (a fixed point).

Now consider the opposite extreme case, where
wavepackets $|a\rangle$ and $|b\rangle$ lie on 
different periodic orbits of the same classical
action (and not related by any symmetry). Then the
two local densities of states $S_a(E)$ and
$S_b(E)$ share the same linear envelope (coming
from short time dynamics), but have completely
uncorrelated long-time fluctuations:
\begin{eqnarray} I_{na}&=&r_{an} S_{\rm lin}(E_n)
\nonumber \\ I_{nb}&=&r_{bn} S_{\rm lin}(E_n)
\end{eqnarray} with $r_{an}$, $r_{bn}$
uncorrelated chi-squared variables with mean $1/N$
(see Eq.~\ref{ran}). Then 
\begin{eqnarray} P_{aa}&=& P_{bb} ={ F \over N}
g(\lambda) \nonumber \\ P_{ab}&=& { 1\over N}
g(\lambda) \nonumber \\ C_{ab} &=& {1 \over F} \,.
\end{eqnarray} The correlation in this case is of
order unity but still less than one.

Finally, choose the two wavepackets $|a\rangle$ and
$|b\rangle$ lying on the same orbit but not exact
time-iterates of one another. We can think of
wavepacket $|b\rangle$ as having a part composed
of $|a\rangle$ and its short-time iterates and
another part which is statistically independent of
$|a\rangle$ although it lies on the same orbit.
The fraction of $|b\rangle$ which is correlated
with
$|a\rangle$ is given by a (normalized) sum of
squared overlaps of $|b\rangle$ with
$|a\rangle$ and its iterates:
\begin{equation}
\label{fraction} { \sum_s |\langle b |U^s_{\rm
lin}|a \rangle|^2 \over
\sum_s |\langle a |U^s_{\rm lin}|a \rangle|^2}
\equiv { g_{ab}(\lambda) \over g(\lambda) } \,,
\end{equation} where Eq.~\ref{fraction} serves as
the definition of $g_{ab}(\lambda)$. In both
numerator and denominator linearized evolution is
used, so
\begin{equation}
\label{gabdef} g_{ab}(\lambda) = \sum_s {1 \over
\cosh (\lambda (s+z))}\,,
\end{equation} where $|b\rangle$ is related to
some exact time-iterate of $|a\rangle$ by a
stretch of
$e^{\lambda z}$ of the Gaussian along the unstable
manifold.

We then have a sum of two contributions:
\begin{eqnarray} P_{ab}  & = & { F \over N }
g(\lambda) \left[{g_{ab}(\lambda) \over g(\lambda)}
\right] + {1 \over N} g(\lambda) \left [ 1 -
{g_{ab}(\lambda) \over g(\lambda)}
\right ]  \nonumber \\ & = & { F-1  \over N }
g_{ab}(\lambda)  + {1 \over N} g(\lambda) \,,
\end{eqnarray} giving 
\begin{equation}
\label{cab} C_{ab} = \left [ 1 - { 1 \over F }
\right ] { g_{ab}(\lambda)
 \over g(\lambda)} + {1 \over F} \,.
\end{equation} Now as the exponent $\lambda$
becomes small, {\it any} optimally
oriented wavepacket $|b\rangle$
lying on the periodic orbit begins to look more
and more like an iterate of any other wavepacket
$|a\rangle$. In that limit,
\begin{equation}
\lim_{\lambda \to 0} { g_{ab}(\lambda) \over
g(\lambda)} = 1 \,,
\end{equation} and so the correlation $C_{ab} \to
1$. So the key result is that when scarring is
strong ($\lambda \to 0$), the spectra of all
optimal wavepackets centered on the periodic orbit
in question are virtually identical, making
unnecessary any averaging over width or position
along the orbit:
\begin{equation}
\langle n | \rho | n \rangle \to | \langle
a_\sigma | n \rangle |^2 \,,
\end{equation} for an arbitrary $|a_\sigma\rangle$
along the periodic orbit. Therefore, in this limit
{\it any} wavepacket individually provides a
universal measure of scarring intensities,
obviating the need to construct tubes and other
averaging devices. 

We now proceed to examine quantitatively the
behavior of $C_{ab}$ in Eq.~\ref{cab}. First,
however, we will introduce a model ensemble of
systems which will allow us to test numerically
this quantitative prediction and others obtained
in the following sections.

\subsection{Ensemble averaging over hard chaotic
systems}

\label{kickbaker}

The classical area-preserving map we will use for
our `numerical experiments' is defined on the unit
square $(q,p) \in [0,1] \times [0,1]$, and
consists of two parts. The first step is a
three-strip generalized baker's
map\cite{baker,sscar} with strip widths
$w_0+w_1+w_2=1$. Each vertical strip $i$ of width
$w_i<1$ and height $1$ is stretched horizontally
by a factor of $1/w_i$ and compressed vertically
by a factor of $w_i$ to make it into a horizontal
strip of height $w_i$ and width $1$. The three
strips are then stacked on top of each other (left
becoming bottom and right becoming top) to
reconstruct the unit square. Defining 
$s_i=\sum_{j<i} w_j$ to be the left edge of strip
$i$, we have
\begin{eqnarray}
\label{step1} q'&=&(q-s_i)/w_i \nonumber \\
p'&=&s_i+p w_i \,,
\end{eqnarray} where the initial position $q$ lies
in the $i$-th strip, i.e.
$s_i \le q<s_{i+1}$. The second and final step is
a kicked map\cite{kickmap}
 implemented in the left and right strips of the
square, leaving the middle strip undisturbed:
\begin{eqnarray}
\label{step2} p'' & = & p' - V_{i'}'({q'-s_{i'}
\over w_{i'}}) \, {\rm mod} \, 1
\nonumber \\ q'' & = & s_{i'}+
[(q'-s_{i'})+p''w_{i'} \, {\rm mod} \, w_{i'}] \,.
\end{eqnarray} Here $i'$ denotes the number of the
strip ($0$ or $2$) containing $q'$. The entire
mapping Eqs.~\ref{step1},~\ref{step2} is now
iterated.

The convenience of this two-step model lies in the
fact that for any choice of kick potentials $V_0$
and $V_2$ acting on the left and right strips,
respectively, the middle strip experiences only
baker-like horizontal stretching and vertical
shrinking. Thus, there is always a fixed point of
the system in the middle strip, with coordinates
$q_f=p_f={w_0 \over w_0+w_2}$, and stretching
exponent $\lambda = |\log w_1|$. Furthermore, the
stable and unstable manifolds of this fixed point
are always locally vertical and horizontal, 
respectively, consistent with our canonical form
Eq.~\ref{eqmo}. The kicked maps acting on the left
and right strips serve to provide parameters which
can be easily varied to produce ensemble averaging
over the details of the nonlinear long-time
recurrences without affecting the local dynamics
around the periodic orbit which is being
studied.\footnote{For a billiard system, the
analogous procedure would be to take a given short
periodic orbit and then create an ensemble of
systems by deforming the boundary in such a way
that the original orbit is unaffected.} We choose
kick potentials
\begin{equation} V_{0,2}(x)=-{1 \over 2} x^2  + {
K_{0,2} \over (2 \pi)^2 } \sin 2 \pi x\,,
\end{equation} with $K_{0,2}$ arbitrary
parameters. The condition $|K_{0,2}|<1$ is
sufficient to ensure hard chaos, without regular
regions\cite{ltsc}. More general kicking
potentials could of course have been used, but we
find that the two parameters $K_{0,2}$ provide a
sufficiently large ensemble for our purposes.

The system is quantized in a straightforward and
conventional way, by multiplying the unitary
matrices implementing baker's map and kicked map
dynamics
\cite{baker,kickmap}.

\subsection{Numerical tests}

We proceed to test the density of states
correlations $C_{ab}$ (Eq.~\ref{cab}) for the
fixed point orbit of the map introduced above. The
wavepacket $|a\rangle$ with horizontal width
$\sigma \sqrt \hbar$ and vertical width $\sqrt
\hbar/\sigma$ is placed on the fixed point. We
then define a family of wavepackets $|b(z)\rangle$
of (horizontal) widths $\sigma e^{\lambda z} \sqrt
\hbar$. Notice that for integer $z$,
$|b(z)\rangle$ is an exact iterate of $|a\rangle$
(in the linear approximation), and thus in that
approximation the densities of states are
identical and the correlation $C_{ab}=1$. The
differences are expected to be greatest at
half-integer $z$ where $|b\rangle$ is most unlike
any iterate of
$|a\rangle$. The correlation $C_{ab}$ is now
plotted as a function of $z$ for $0 \le z\le 1$ in
Fig.~\ref{cabfig}. Two sets of data are given,
differing in the stability exponent of the
periodic orbit (which is easily adjusted by varying the
middle strip width $w_1$). The two values used
were $\lambda=\log 5$ (upper curve) and
$\lambda=\log 10$ (lower curve). In each case, the
numerical data comes from an ensemble average over
systems of size $N=1/h=200$. The errorbars shown 
in the figure are statistical, and do not reflect
finite-size effects. The theoretical curves are
obtained from Eq.~\ref{cab} and require only the
single parameter $\lambda$. The agreement between
theory and data is quite good; furthermore we see
just how large the correlations are even for not
very small exponents. Thus, for an orbit with a
stretching factor of $10$ per iteration ($\lambda
= \log 10$), the correlation does not go below
$0.95$ even for the maximally unrelated
wavepackets ($z=1/2$).

In Fig.~\ref{cabfig2} we plot this minimum
correlation $C_{ab}(z=1/2)$ on the vertical axis,
versus the scarring enhancement factor
$g(\lambda)$ on the horizontal axis. Four data
points are used, corresponding (from left to right)
to stretching factors $e^{\lambda}=20$, $10$, $5$,
$2.5$. Both the expected average enhancement $g$
and the inter-wavepacket correlation
$C_{ab}(z=1/2)$ are uniquely given theoretically
as functions of the exponent
$\lambda$. Again, the data agrees very well with
the theoretical predictions. We see that a
scarring enhancement factor of $2$ (corresponding
to stretching exponent  $\lambda \approx \log 5$)
is associated with a {\it minimum} correlation of
$0.99$ between {\it the least correlated}
wavepackets on that orbit. Strong scarring thus
automatically eliminates the ambiguity in measuring the strength
of Gaussian wavepacket scarring.

The question then becomes whether it is possible
in any way to take advantage of our knowledge of
the orbit and its invariant manifolds to produce a
scarring test state that would do better than a
single Gaussian wavepacket. Indeed, this is
possible, and what is necessary is to use {\it
coherent} quantum superpositions of test
states\cite{saraceno,borondo,tomsovic} instead of
the density matrix approach investigated in this
section.

\section{Coherent wavepacket sums: enhanced
scarring}
\label{coh}

\subsection{Theory}

As suggested already in \cite{nlscar}, we can
construct a ``linearized eigenstate" $|\psi
\rangle$ as a normalized coherent sum of Gaussian
wavepackets centered on a periodic orbit. For a
fixed point orbit, we write
\begin{equation}
\label{cohdef}
\Psi={\cal N} \int dt \, e^{-t^2/T_0^2} \,
|a_{\sigma e^{\lambda t}} \rangle
\end{equation} (see Eq.~\ref{rhodef}). $T_0$ is a
linearized dynamics time cutoff as defined in
Eq.~\ref{t0def}, and the normalization constant
${\cal N}$ ensures $\langle \Psi | \Psi \rangle
=1$. Just as was done for the density matrix in
the preceding section, Eq.~\ref{cohdef} can be
generalized  in a straightforward way to
higher-period orbits and to continuous time.
However, to make the presentation more transparent
the examples here and in the following section are
restricted to the case of a fixed point, the
generalizations being left to Section~\ref{hiper}.

If the dynamics
away from the periodic orbit were exactly linear,
we could take the cutoff $T_0$ to infinity and
obtain a stationary state with quasi-energy $\phi$
(phase $e^{-i\phi}$, see Eq.~\ref{wpevol}): hence
the name ``linearized eigenstate". In reality, a
finite cutoff is necessary because the ratio of
the size of the linearizable region of phase space
$A$ to $\hbar$ is finite. However, if this ratio
is large (as it will always be in the
semiclassical limit $\hbar \to 0$), most of
  the state $\Psi$ maps to itself
under Eq.~\ref{wpevol}, producing a  large
autocorrelation function at short times. In the
case of Gaussian wavepacket scarring, the extent
to which the short-time  return probability
differs from unity (and thus the extent to which
perfect localization fails to be achieved) is
determined by the instability of the orbit (i.e.
by the amount by which $\lambda$ is different form
zero). As measured using the improved test state
$\Psi$, the absence of complete localization is
given by the failure of the {\it linearized}
dynamics at long times. 

The test state $\Psi$ lives not only at the
periodic point, but also along the invariant
manifolds.  Its autocorrelation
$\langle\Psi\vert\Psi(t)\rangle$ decays only on
the order of the  {\it log-time}
$T_0 \sim \lambda^{-1} \log {A/\hbar}$, as we
show
explicitly below in Eq.~\ref{apsi2}.  This makes
$\Psi$ a much sharper measure of the scar
character of an eigenstate, and for small $\hbar$
we expect to see much stronger localization as
measured by $\Psi$ than by an
 individual wavepacket
$|a_\sigma\rangle$. We will now proceed to show
this explicitly. The construction of $\Psi$ is
extremely simple, requiring only one piece of
information beyond what we already needed for the
single wavepacket, namely the (approximate) size
of the region in which the dynamics is
linearizable. No knowledge of long-time dynamics,
nonlinear recurrences, or any other periodic
orbits is needed.

We also notice that in the strong scarring limit
$\lambda \to 0$, we could just as well have used
only the linearized time iterates of $|a \rangle$
(rather than wavepackets of all widths) as in
Eq.~\ref{cohdef} to construct the hyperbolic test
state: 
\begin{equation}
\label{cohdyn}
\Psi_{\rm dyn} \sim \sum_t \, e^{-t^2/T_0^2} \,
|a_{\sigma e^{\lambda t}} \rangle =
\sum_t \, e^{-t^2/T_0^2}e^{i \phi t} \, |a_{\rm
lin}(t) \rangle
\end{equation} (cf. Eq.~\ref{rhodyn}). This form
makes manifest the close connection   between the
construction of the scarring test state and the
linearized classical dynamics (Eq.~\ref{eqmo}). It
also makes almost trivial the generalization to 
higher period orbits and to continuous time (see
Eq.~\ref{rhodyn} and also the fuller discussion in
Section~\ref{hiper}). The main disadvantage of the
form $\rho_{\rm dyn}$ (as compared to $\rho$) is
that the former requires the arbitrary choice of
initial width $\sigma$. However, as we have seen
in the previous section, this choice of starting
wavepacket has no effect on any measured
quantities in the
$\lambda \to 0$ limit (where replacement of the
integral by a sum is justified).

We begin as in Section~\ref{gausscar} by
evaluating the short-time autocorrelation function
\begin{equation} A_{\rm lin}^\Psi (t)\equiv
\langle \Psi | U_{\rm lin}(t) | \Psi \rangle \,.
\end{equation} A straightforward calculation using
Eqs.~\ref{wpevol},~\ref{shortcorr} gives:
\begin{equation}
\label{apsi} A_{\rm lin}^\Psi (t) = Q e^{-i \phi t}
\int dy \, { e^{-(t-{y \over \lambda})^2/T_0^2}
\over 
\sqrt{\cosh y}} \,.
\end{equation} The overall normalization constant
$Q$ can be fixed by requiring
$A_{\rm lin}^\Psi (0) = \langle \Psi | \Psi
\rangle =1$. The integration variable $y$ is a
time variable scaled by $\lambda$ to make it
dimensionless. In the limit $T_0 \lambda \gg 1$,
i.e. 
$\log {A \over \hbar} \gg 1$, the exponential
simplifies and we obtain
\begin{eqnarray}
\label{apsi2} A_{\rm lin}^\Psi (t) & =&  Q e^{-i
\phi t}
 \int dy \, { e^{-t^2/T_0^2} \over \sqrt{\cosh y}}
\nonumber \\ & = & e^{-i \phi t} e^{-t^2/T_0^2} \,.
\end{eqnarray} 
Now we see explicitly that the
decay rate of our test state $\Psi$ is indeed
given by the log-time scale $T_0$. Of course for
the {\it linear} autocorrelation function to be a
good measure of the {\it total}  return amplitude
$A^\Psi (t)$, even for times $t$ less than $T_0$,
the state $\Psi$ must be well contained inside the
linear region. This can be done by adjusting the
constant $c$ in Eq.~\ref{t0def}: numerically we
will see below that good quantitative agreement
with Eq.~\ref{apsi} is obtained for $c \approx
0.6$. In any case, the precise value of this
constant does not affect any of the important
scaling arguments which will follow.

As in Section~\ref{gausscar}, the inverse
participation ratio has an enhancement factor
associated with the short time recurrences:
\begin{eqnarray} {\rm IPR}_\Psi & \equiv & { <
|\langle n|\Psi\rangle|^4 >
\over < |\langle n|\Psi\rangle |^2 >^2 }  \\
\label{iprpsi} &=& F \sum_t |A_{\rm lin}^\Psi
(t)|^2 \\
\label{iprpsi2} & \approx &F \sqrt \pi T_0 \,,
\end{eqnarray} where in the last line the limiting
form Eq.~\ref{apsi2} has been used, and $T_0$
taken to be large. In the Gaussian wavepacket case,
the IPR scaled with the orbit decay time
$\lambda^{-1}$; here it scales with the log-time
$\sim \lambda^{-1} |\log \hbar| \gg \lambda^{-1}$.
This makes the coherent test state a factor
of $T_0 \lambda/\sqrt \pi \sim |\log \hbar|$
times better than any of the 
single Gaussian test states.

We can also look at the spectral envelope
$S^\Psi_{\rm lin}$ which is the fourier transform
of the short-time autocorrelation function (see
Eq.~\ref{smspectrum}):
\begin{eqnarray}
\label{slinpsi} S_{\rm lin}^\Psi(E)& =&\sum_t
A^\Psi_{\rm lin}(t) e^{iEt} \\
\label{slinpsi2} & \approx & \sqrt{2 \pi} T_0
e^{-(E-\phi)^2T_0^2/2} \,,
\end{eqnarray} where again in the last line the
limiting ($\hbar \to 0$) form Eq.~\ref{apsi2} has
been used. The energy envelope is centered at
$E=\phi$, just like the smoothed single-wavepacket
local density of states, but the the peak is both
narrower and taller by a factor scaling as $|\log
\hbar|$.

\subsection{Numerical Tests}

We now check the results
obtained in this section, using again the ensemble
of kicked-baker systems introduced in
Section~\ref{kickbaker}. The short periodic  orbit
will again be the fixed point of the middle
baker's strip, with exponent $\lambda$ set by the
logarithm of the width of this strip. We begin by
looking at the smoothed local densities of states
$S_{\rm lin}^\Psi$ and $S_{\rm lin}^a$, for the
universal test state $|\Psi\rangle$ and the simple
Gaussian
$|a\rangle$, respectively. The width $\sigma$ of
the starting Gaussian is set to $\sqrt w_1$, so
that the aspect ratio of the Gaussian is equal to
the aspect ratio of the rectangular middle strip
in which the classical dynamics is linearizable.
The wavepacket can then expand the same number of
steps in either time direction before reaching the
edge of the linear regime. The test state $\Psi$
is constructed using a cutoff set by $c=0.6$ (see
Eq.~\ref{t0def}).

Ensemble averaging is performed over many
kicked-baker systems of the same (reasonably large)
exponent $\lambda=|\log 0.18|$, and of system size
(Hilbert space dimension) $N=1/h=800$. Local
densities of states for
$|a_\sigma \rangle$ and $|\Psi\rangle$ are
ensemble-averaged and smoothed, with the resulting
envelopes plotted in Fig.~\ref{specfig}.
Theoretical curves obtained form
Eqs.~\ref{smspectrum},~\ref{slinpsi} are also
plotted for comparison. Excellent agreement is
observed between the data and the predictions
based on the linearized theory. Furthermore, we
see that the spectral envelope for the hyperbolic
test state $\Psi$ is significantly narrower and
taller than the corresponding envelope for the
Gaussian wavepacket, again in accordance with
prediction. We should note here that the
hyperbolic test state is constructed here with the
very modest log-time cutoff $T_0 = 0.90$. There
are three reasons for the smallness of $T_0$ in
this example: 1) the stretching factor $e^\lambda
\approx 5.6$ is rather large, 2) the system size,
and particularly the size of the linearizable
region, are modest, 3) and finally the free
parameter $c$ has been set at a rather
conservative (small) value. With regard to the
last point, we should note in particular that
increasing the cutoff parameter $c$
(Eq.~\ref{t0def}) will give rise to a sharper
envelope, with larger inverse participation ratio,
though at some cost to the accuracy of the
formulas Eqs.~\ref{iprpsi},~\ref{slinpsi}, etc. In
effect, there is  a tradeoff between keeping the
test state well inside the linear region and thus
being able to obtain with good accuracy its
statistical properties (smaller $c$) versus
maximizing the localization properties of the
hyperbolic test state by allowing it to some
extent to leak out of the linear region (larger
$c$). All of this will become clearer as we go on
to discuss IPR measures for the universal test
states. Of course, none of these
$O(1)$ considerations affect the basic scaling
predictions: namely the height, inverse width, and
IPR of the spectral envelope for $\Psi$ all scale
inversely  with $\lambda$ for small $\lambda$ and
also logarithmically with $1/\hbar$ for small
$\hbar$. In particular, the hyperbolic
(coherent $\Psi$ test state) spectral envelope gets
arbitrarily taller and narrower as
$\hbar \to 0$ for a fixed classical system, while
the corresponding spectral envelope for a single
Gaussian packet 
test state 
remains unchanged.

We next probe the behavior of the
mean ${\rm IPR}_\Psi$ as a function of $N$ and $c$
(with exponent $\lambda$ again set to $|\log
0.18|$). In Fig.~\ref{iprfig}, we plot the IPR
versus cutoff time $T_0$, for five sets of data:
$N=1/h=50$, $100$,
$200$, $400$, and $800$ from bottom to top. For
each value of $N$,
$26$ values are plotted: from left to right
$c=(1.1)^j$, $j=-20 \ldots +5$. The five values at
$T_0=0$ represent simple Gaussian test states ($c
\to 0$). The upper dashed curve represents the
theoretical prediction of Eq.~\ref{iprpsi}, which
should hold for large values of $N$. Good
agreement with the data is obtained for $N \ge
200$, as finite-size effects become less
relevant.\footnote{Notice that because the width
of the central strip here is quite small ($0.18$),
$N=200$ corresponds to a size of only $0.18 \times
200 =36$ for the linearizable region (in units of
$h$).} The six rightmost points on each data curve
represent $c \ge 1$, and some deviation from the
linear theory prediction is expected to start
setting in there. We also note that  at very large
values of $T_0$ (requiring correspondingly larger
values of $\log N$), the  theoretical prediction
converges to the linear  asymptotic form of
Eq.~\ref{iprpsi2} (lower dashed line).

We note for purposes of comparison that the
single-wavepacket scarring strength
${\rm IPR}_a$ is predicted to be
$3.66$ for this value of $\lambda$ (see
Eq.~\ref{scaripr}, noting that a quantum
fluctuation factor
$F=2$ is appropriate for complex eigenstates).
This indeed is close to the value  attained by the
single wavepackets  ($T_0=0$), at least for $N \ge
200$.
\footnote{The deficit in the measured single
Gaussian IPR values ($T_0=0$) for finite $N$, i.e.
the extent to which these fall below the limiting
value of $3.66$, is in close correspondence with
similar deficits in the hyperbolic state IPR's
($T_0 >0$).} We see that IPR values significantly
larger than this can be attained using the
hyperbolic test states, especially for larger
values of $N$.  The data is consistent with the
prediction that the IPR (at fixed value of $c \sim
1$) scales logarithmically with $N$ for large
system size $N$.

Finally, to close this section we present in
Fig.~\ref{husfig} the Husimi phase-space plot for
the hyperbolic test state $\Psi$. In this figure,
$T_0$ has been taken to be very large compared to
$\lambda^{-1}$, i.e. the linearizable regime is
much larger than a unit Planck cell, and also much
larger than the phase-space area shown in the
figure. We note that the phase-space Husimi
picture is universal, and in particular 
independent of the exponent $\lambda$, since a
change in the value of $\lambda$ in
Eq.~\ref{cohdef} can of course always be absorbed
into a redefinition of $T_0$ and the overall
normalization $\cal N$. In other words, in the 
$\log {A \over \hbar} \to \infty$ limit, the state
$\Psi$ depends only on the linear region size
parameter $A/\hbar$, and when we further  look
well inside the area $A$, its structure is
completely free of any parameters.  The phase
space area shown in the figure is $12 \sqrt\hbar
\times 12
\sqrt \hbar$, i.e. it contains $144/2\pi$
Planck-sized cells. We see from the Figure that
$\Psi$ lives at the periodic point in the center
of the plot (the size of the bright region at the
periodic point being set by $h$), and
symmetrically along the linearized stable and
unstable manifolds. This picture will be important
to us for comparison purposes when we study
off-resonance universal test states in the
following section. The analytic expression used to
obtain the density plot in Fig.~\ref{husfig} will
also be given there (Eq.~\ref{hus}).

\section{Off-resonance and off-orbit scarring}
\label{offres}

Going back to the construction of the hyperbolic
test state $\Psi$ in Eq.~\ref{cohdef}, we notice
that there the Gaussian wavepackets are all added
{\it in phase}, giving rise to a preferred energy
$\phi$ which is the same as that for any single
wavepacket considered individually. We may,
however, equally well consider the more general
form
\begin{equation}
\label{cohdefth}
\Psi={\cal N} \int dt \, e^{-t^2/T_0^2} e^{i
\theta t}\, |a_{\sigma e^{\lambda t}} \rangle \,,
\end{equation} where $\theta$ is an arbitrary
phase accumulated per time step. This extra phase
should give rise to a state that prefers to live
at an energy different from the one that exactly
quantizes the periodic orbit (i.e. $E=\phi$). In
turn, this energy shift may be expected to give
rise to phase space structures that lie away from
the invariant manifolds of the periodic orbit, i.e.
above and below the separatrix constructed from
these manifolds\cite{voros}. These intuitive
expectations turn out to be justified, as we now
shall see.

We begin once again by computing the short-time
(linearized) autocorrelation function. The
generalization of Eqs.~\ref{apsi},~\ref{apsi2}
works out to be:
\begin{eqnarray}
\label{apsith} A_{\rm lin}^\Psi (t) &=& Q e^{-i
\phi t}
\int dy \, { e^{-(t-{y \over \lambda})^2/T_0^2-i
\theta(t-{y \over \lambda})}
\over
\sqrt{\cosh y}}  \\ \label{apsi2th} & \approx &
e^{-i (\phi+\theta) t} e^{-t^2/T_0^2} \,.
\end{eqnarray} 
In the last line, the limits
$\lambda \to 0$ and $T_0 \lambda \sim
\log {A \over \hbar} \to \infty$ have been taken,
in complete analogy with Eq.~\ref{apsi2}.

>From these expressions, we proceed to obtain
spectral envelopes and IPR values, exactly as we
did previously for the $\theta=0$ special case. 
It is interesting to note here that in the
asymptotic limit of Eq.~\ref{apsi2th}, the
spectral envelope will be exactly the same as that
obtained previously (Eq.~\ref{slinpsi2}), but
shifted so as to be centered at energy
$E=\phi+\theta$. Because the shape of the envelope
is unaffected by the value of $\theta$ in this
limit, the IPR is still asymptotically given by
our previous formula, Eq.~\ref{iprpsi2}. It is
important to note, however, that for realistic
values of $\log {A \over \hbar}$, this asymptotic
regime may not be attained, and the more general
formulas Eqs.~\ref{iprpsi},~\ref{slinpsi},
\ref{apsith} should be used instead.

In Fig.~\ref{specfigth}, we plot (numerically
obtained) smoothed spectral envelopes for the
hyperbolic states $\Psi$, for the same ensemble of
systems as was used in producing
Fig.~\ref{specfig}. The envelopes correspond to
$\theta=0$ (the tallest envelope, already seen
previously in Fig.~\ref{specfig}), through
$\theta=-2\pi$, moving to the left in steps of
$2\pi/20$. Even though the asymptotic form of
Eq.~\ref{apsi2th} predicts all the envelopes
should have the same shape, being merely shifted
to the left by angle $\theta$, in reality we see
this is not quite the case for finite values of
$\log(A/\hbar)$. Because of the finite
linearizable volume of phase space, the
off-resonance hyperbolic states have significantly
less well-defined envelopes compared to the
$\theta=0$ state. That is because the asymptotic
form assumes most of the autocorrelation
function comes from long-time overlaps of
wavepackets with very different widths. At finite
system sizes, a very important correction is the
partial self-cancellation in $\Psi$ coming from 
wavepackets of comparable widths being added
together with very different phases. This
correction is, of course, taken into account in
the more general form of Eq.~\ref{apsith}, which
does in fact predict less sharp envelopes (and
consequently lower IPR's) for the off-resonance
states. The  important point to notice here,
however, is the presence of a very significant
localization effect even for $|\theta| > \lambda$,
i.e. at energies well outside the resonance of the
original Gaussian wavepacket. States at such
energies are not strongly scarred according to the
original (Gaussian wavepacket) definition, nor are
they particularly enhanced along the stable and
unstable manifolds of the orbit, as measured by
the on-resonance hyperbolic test state.  However,
such states do have enhanced intensity relative to
the off-resonance hyperbolic states $\Psi(\theta
\ne 0)$, which live in hyperbolic regions on
either side of the separatrix (see
Fig.~\ref{husfigth} later in this section).
Although the enhancement factors for such states
are quite modest for the parameters chosen ($\sim
2$ over RMT), they will of course grow with
$\lambda^{-1}$ and with $\log{A/\hbar}$, as we
discussed earlier. In particular, in the high
energy limit ($\hbar \to \infty$) of a given
classical system, scarring by the off-resonance
states is expected to be equally strong compared
to  the on-resonance ($\theta=0$) form of scarring.

To demonstrate the preceding assertion, we plot in
Fig.~\ref{iprthfig} the (theoretically computed)
IPR as function of $T_0$, for several values of
the off-resonance angle $\theta$. To facilitate
comparison with Fig.~\ref{iprfig}, we again choose
stretching exponent $\lambda=|\log 0.18|$. The six
curves represent (from top to bottom at $T_0=1$)
values of $\theta$ from
$0$ to $\pi$, in steps of $\pi/5$. The top curve
has thus already appeared previously in
Fig.~\ref{iprfig}. The lowest curve will later be
compared with data in Fig.~\ref{iprpifig}. The
asymptotic form of Eq.~\ref{iprpsi2}, to which all
these curves converge at large $T_0$ (large $|\log
\hbar|$), is shown in the figure as a dashed line.

An interesting point to notice here is that for
sufficiently large $\theta$, the IPR can drop
at moderate $T_0$
from its single-wavepacket value (at 
$T_0=0$), before eventually recovering at larger $T_0$ (this
is a result of the
self-cancellation effect alluded to earlier).
However,  for (exponentially)
large systems, scarring is equally strong for the
different values of the off-resonance parameter $\theta$.
Intuitively, for a large system (or small $\hbar$),
the size of the linearizable region in which the
test state $\Psi$ is constructed is very large compared
to the size ($\sim h$) of a single wavepacket. Thus,
a given wavepacket used into the construction of $\Psi$
in Eq.~\ref{cohdefth} has very little overlap with most
of the other wavepackets in the sum. For this reason,
the cancellation
effect arising from similar wavepackets being added together
with different phases (for non-zero $\theta$)
becomes insignificant
in the semiclassical limit.

To make contact once again with the data, we
choose $\theta=\pi$, and in Fig.~\ref{iprpifig} do
for this case what we did for $\theta=0$ earlier
in Fig.~\ref{iprfig}. Again, data for $N=50$,
$100$, $200$, $400$, and
$800$ is plotted, and the same range of $c$-values
is used. Only for $N \ge 200$ do we see the
recovery towards larger IPR values that we expect
>from the theoretical curve (dashed). Recall that
$3.66$ is the single-wavepacket IPR value.

In Fig.~\ref{iprvsth}, the system size $N=200$ is
fixed, with cutoff parameter $c=0.6$, and the IPR
is plotted versus the off-resonance angle
$\theta$, ranging here from $0$ to $2\pi$.
Agreement between theory and data is in this case
(surprisingly) good. We see for this value of $N$
that by $\theta=2\pi$, the IPR values are already
approaching the RMT value of 2, indicating almost
no localization.

Finally, we return to the Husimi representation of
$\Psi$, which we began to discuss already in the
previous section. Husimi representations
of the off center scarring were discussed in
\cite{voros}, where the inverted oscillator
eigenstates were probed with Husimi states. 
Here, we have shown how to generate test states
which are sensitive to off-center scars developing
in chaotic systems.
For
reference, the expression describing the Husimi
intensities is:
\begin{eqnarray} H^\Psi_{q_0,p_0}&=& |\langle
g_{q_0,p_0} |\psi \rangle|^2 \nonumber \\
\label{hus} &=& e^{-q_0^2/\hbar}
\left |\int dy {e^{-y^2/\lambda^2 T^2}
e^{iy\theta/\lambda}  e^{(q_0-ip_0)^2 e^y
\over 4 \hbar \cosh y} \over \sqrt{\cosh y}}
\right |^2 \,.
\end{eqnarray} Here $g_{q_0,p_0}$ is a phase-space
Gaussian centered at phase-space point
$(q_0,p_0)$ of width $\sqrt\hbar$ in both position
and momentum:
\begin{equation}
\label{gaus} g_{q_0,p_0}(q)=(\pi
\hbar)^{-1/4}e^{-(q-q_0)^2/2\hbar+ip_0(q-q_0)/\hbar}
\,.
\end{equation} Recall that we are working in a
coordinate system where the fixed point is located
at the origin $(q,p)=(0,0)$, and the linearized
invariant manifolds  are horizontal and vertical
(Eq.~\ref{eqmo}). As the linearizable region 
becomes large compared to the phase space region
of interest ($T_0 \to \infty$), the Husimi density
of Eq.~\ref{hus} depends on only three
dimensionless numbers: an off-resonance parameter
$\theta/\lambda$ and the phase space coordinates
$q_0/\sqrt\hbar$, $p_0/\sqrt\hbar$.

Previously, in Fig.~\ref{husfig}, we have seen the
Husimi density of $\Psi$ for $\theta/\lambda=0$
plotted in the square phase space area $-6 \le
q_0/\sqrt\hbar, p_0/\sqrt\hbar \le +6$. In
Fig.~\ref{husfigth} we present the analogous
picture for (a) $\theta/\lambda=0.8$, and (b)
$\theta/\lambda=2.5$. As the energy goes further
off resonance, the state $\Psi$ moves away from
the periodic orbit and its invariant manifolds,
and shifts into two of the quadrants separated by
the manifolds ($qp >0$ or $qp<0$, depending on the
sign of $\theta$). For large $|\theta|/\lambda$,
narrow hyperbolic regions in phase space are
accessed, lying further and further from the
periodic orbit itself. For $\theta/\lambda \sim
\pm 1$, $\Psi$ lives near the hyperbolas 
$qp \sim \pm \hbar$ surrounding the periodic
point. This regime corresponds to a spectral
envelope for $\Psi$ which is centered at energy
$\lambda$ away from the EBK energy, i.e. at the
edge of the single-Gaussian spectral envelope.
Eigenstates having strong overlaps with $\Psi$ are
now barely scarred at the periodic point itself.
If we go further into the regime $|\theta|/\lambda
\gg 1$, the spectral envelope of $\Psi$ is now
centered at an energy which is outside the
envelope  of the single wavepacket. Then, states
overlapping such a test state $\Psi$ will have
stronger than expected intensity on hyperbolic
regions surrounding the periodic orbit, but will
not be scarred at all (and may even be
antiscarred) on the orbit itself.

The connection discussed above between an energy
shift away from the EBK value and hyperbolic
phase-space structures may be understood very
simply by considering the evolution of off-center
Gaussians $g_{q_0,p_0}$ of Eq.~\ref{gaus}. Such a
Gaussian is not an optimal test state for
measuring scarring, because the autocorrelation
function decays quite rapidly, especially for
$q_0^2+p_0^2 \gg \hbar$ (to be contrasted with
$\Psi$, which lives along the {\it entire}
hyperbolic region, and thus has much larger
self-overlaps at short times). However, the phase
information in the autocorrelation function for
$g_{q_0,p_0}$ is quite relevant:
\begin{equation} A_{\rm lin}(t)={e^{-i \phi t}
e^{-{q_0^2+p_0^2 \over \hbar} {\sinh^2{\lambda
t/2} \over \cosh{\lambda t}}}} e^{-{iqp \over
\hbar} \tanh{\lambda t}} 
\end{equation} (compare with Eq.~\ref{shortcorr}).
Upon fourier transforming this to obtain a
spectrum, we obtain an expression for the optimal
energy as a function of phase-space location:
\begin{equation} E-\phi \approx \lambda {qp \over
\hbar} \,.
\end{equation} Thus the off-resonance parameter
$\theta$ of our test state $\Psi$ is then expected
to be associated with phase-space hyperbolas:
\begin{equation} {\theta \over \lambda } \approx 
{qp \over \hbar} \,.
\end{equation}

\section{Extension to higher-period orbits and
continuous time}
\label{hiper}

\subsection{Longer orbits of maps}

The analysis of the previous two sections has
focused on fixed-point periodic orbits, but it
generalizes in a straightforward way to longer
orbits and to continuous time. We shall see below
that the benefits of using universal scar measures
instead of simple Gaussians become even greater
when longer orbits are considered.

Consider again a periodic orbit (of a map) of
period $T_P$, with periodic points $x_p$ ($p=0
\ldots T_P-1$), as in Eq.~\ref{orbsum}. Let
$-\phi$ be the phase accumulated over one full
iteration of the orbit, and $\lambda$ the
corresponding stability exponent. The short-time
autocorrelation function for a Gaussian $|a
\rangle$ centered at any of the periodic points is
a generalization of Eq.~\ref{shortcorr}:
\begin{equation}
\label{ap2} A_{\rm lin}(t)={e^{-i \phi t/T_P}
\over \sqrt{\cosh{\lambda t/T_P}}}
\delta_{(t \, {\rm mod} \, T_P),\, 0} \,.
\end{equation} The IPR as function of $\lambda$ is
then the same as for a period-one orbit (because
it is given simply by the sum of the short-time
return probabilities, Eq.~\ref{iprsum}), while the
short-time spectral envelope $S_{\rm lin}$ has
$T_P$ peaks in the  quasienergy domain $[0,2\pi]$,
of height scaling as $\lambda^{-1}$, and width
scaling as $\lambda/T_P$. The peak energies are of
course those that `quantize' the orbit:
$T_P E=\phi \, {\rm mod} \, 2\pi$, or
\begin{equation} E_k= {\phi + 2\pi k \over T_P} \;
\; , \; \; k=0 \ldots T_P-1 \,.
\end{equation} We notice that both the maximum
scarring strength and the IPR can be large only
for small $\lambda$, which becomes difficult to
achieve for the longer orbits $T_P >1$ (normally
$\lambda$ grows linearly with $T_P$).

We now proceed to construct the universal test
state $\Psi$ for such  an orbit, having made a
choice of quantization energy $E_k$:
{\begin{eqnarray}
\label{hiperpsi}
\Psi & = &{\cal N} \sum_p \int dt e^{-t^2/T_P^2
T_0^2} \nonumber \\ & \times & e^{i (E_k +
\theta/T_P)p +i\theta t/T_P -i \phi_p }
|a_{x_p,\sigma f_p e^{\lambda t/T_P}} \rangle \,.
\end{eqnarray} Here $f_p$ is a stretching factor,
and $\phi_p$ is a phase, both defined by
\begin{equation} U_{\rm lin}^p |a_{x_0,\sigma}
\rangle = e^{-i \phi_p}|a_{x_p,\sigma f_p
e^{\lambda p/T_P}}\rangle  \,.
\end{equation}
$f_p$ and $\phi_p$ take into account the fact that
stretching and phase accumulation along the orbit
may both be non-uniform; of course, 
$f_{T_P}=1$ and $\phi_{T_P}=\phi$. The factors
$f_p$ are of order unity and thus not very
important in the semiclassical limit
$\log {A/\hbar} \to \infty$ when the linearizable
region is very large; on the other hand, the
phases $\phi_p$ are crucial for getting
constructive interference. The parameter $T_0$ is
defined as before (Eq.~\ref{t0def}), using the
area of the linearizable region around periodic
point $x_0$, and $\theta/\lambda$ is an
off-resonance parameter, as discussed previously.

The short-time autocorrelation function of the
state $\Psi$ then has the same form as what we
found previously for the special case $T_P=1$
(Eqs.~\ref{apsi},
\ref{apsi2},~\ref{apsith},~\ref{apsi2th}),
replacing
\begin{equation} T_0 \to T_0 T_P \;,\;\lambda \to
\lambda /T_P \;,\; 
\theta \to \theta /T_P  \;,\; \phi \to E_k
\end{equation} throughout. Note that $\lambda$ and
$\theta$ are defined as stretching  exponent and
phase {\it per orbit period} rather than per time
step, and likewise
$T_0$ is the log-time measured in units of the
orbit period. Eqs.~\ref{slinpsi},~\ref{slinpsi2}
describing the shape of the linear spectral
envelope, and Eqs.~\ref{iprpsi},~\ref{iprpsi2} for
the IPR of the universal test state $\Psi$ undergo
the same simple modifications and are then
applicable to the case of $T_P>1$.

Let us compare these results with ordinary (single
Gaussian) scarring for $T_P>1$ as well with the
scarring of a fixed point, which we focused on in
the previous two sections.

For general $T_P \ge 1$, the autocorrelation
function of a Gaussian wavepacket
$|a_\sigma\rangle$ has $O(\lambda^{-1})$ strong
recurrences spaced $T_P$ steps apart, and thus
stretching over a total time scale  of order $T_P
\lambda^{-1}$. In the quasienergy domain, this
leads to a set of $T_P$ evenly spaced spectral
envelopes, each with width, height, and IPR
scaling  (for small $\lambda$) as:
\begin{eqnarray} w_a & \sim & \lambda / T_P
\nonumber \\ h_a & \sim & \lambda^{-1} \nonumber \\
{\rm IPR}_a & \sim & \lambda^{-1} \nonumber \\ (
T_P \;\; {\rm envelopes,}  & & {\rm centered} \;\;
{\rm  at}
\;\; {\rm all} \;\; E_k) \,.
\end{eqnarray} Although the width scales with the
exponent {\it per time step}, due to the presence
of $T_P$ of these envelopes, the maximum expected
scarring strength and the IPR both scale only with
the total exponent {\it per iteration of the
entire orbit}, and thus are expected to deviate
less and less from perfect ergodicity as longer
orbits are considered.

Let us repeat the same analysis for the state
$\Psi$, which takes  properly into account all of
the linearized dynamics around the periodic orbit.
The short-time autocorrelation function does not
decay until a time of order
$T_0 T_P \sim T_P \lambda^{-1} \log {A /\hbar}$.
This produces a {\it single} peak centered at
quasienergy $E_k$ (or shifted by an off-resonance
phase $\theta/T_P$). The width, height, and IPR
scale as
\begin{eqnarray} w_a & \sim & \lambda / (T_P  \log
{A /\hbar}) \nonumber \\ {\rm IPR} \sim h_a & \sim
& (T_P \log {A /\hbar}) / \lambda
\nonumber  \\ ( 1 \;\; {\rm envelope,} & & {\rm
centered} \;\; {\rm  at} \;\; {\rm  some} \;\;
E_k+{\theta \over T_P}) \,.
\end{eqnarray} Apart from the logarithmic
enhancement which leads to more and more deviation
from RMT in the $\hbar \to 0$ limit, we also
notice that only one peak is present (a choice of
quantization energy $E_k$ having been made), and
all quantities now depend only on the ratio
$\lambda/T_P$, the stability exponent {\it per
time step}. This measure of scarring therefore
allows us to see strong effects even for longer
periodic orbits, as long as the stretching per
time step remains moderate.

If the total exponent $\lambda$ (and not just
$\lambda/T_P$) is small, we can equivalently use
the linearized dynamics to generate our universal
test state (compare
Eqs.~\ref{rhodyn},~\ref{cohdyn}). Eq.~\ref{cohdyn}
generalizes easily to
\begin{equation}
\label{hiperdyn}
\Psi_{\rm dyn} \sim \sum_t \, e^{-t^2/T_P T_0^2} \,
e^{i(E_k+\theta/T_P)t} |a_{\rm lin}(t) \rangle\,,
\end{equation} where $|a \rangle$ is a Gaussian
wavepacket of width $\sigma$  centered at any
point along the periodic orbit. Eq.~\ref{hiperpsi}
can be thought of as an averaging of
Eq.~\ref{hiperdyn} over the initial width $\sigma$
from some $\sigma_0$ to $\sigma_0 e^\lambda$. For
small
$\lambda$, the averaging procedure is unnecessary,
all the states being essentially identical, and
the simple expression of Eq.~\ref{hiperdyn} well
describes the universal test state for any choice
of $\sigma$.

\subsection{Hamiltonian systems}

The entire analysis can be applied also to
Hamiltonian systems in an essentially unchanged
form. Let $T_P$ again be the period of the orbit
(now measured in real time units rather than in
time steps), and let $\lambda$ and $\phi$ still be
the exponent and phase per one iteration of the
orbit. Then an optimal Gaussian centered anywhere
on the orbit and aligned along the stable and
unstable manifolds (with width $\sigma$ along the
unstable manifold) has a linear autocorrelation
function given by a sum over iterations of the
orbit:
\begin{equation}
\label{hamwpkt} A_{\rm lin}(t)=\sum_n w(t-nT_P)  
{e^{-i \phi n} \over \sqrt{\cosh{\lambda n}}}
\,.
\end{equation} The very-short-time window function
$w(t)$ describes the self-overlap of the
wavepacket as it intersects itself once every
period; it is associated with the nonzero width
$\sigma_x$ of the wavepacket in the direction of
the orbit, and has a scale 
$\Delta t \sim \sigma_x / v \sim \sigma_x /\sqrt
E$. (In the last equality, the particle mass has
been assumed to be unity, as it will be
throughout.) Fourier transforming, we obtain a
very wide envelope (of width
$\Delta E \sim \hbar \sqrt E /\sigma_x$)
associated with the energy uncertainty of the
wavepacket itself. Multiplying this are the
scarring envelopes, centered at the quantizing
energies, of width\footnote{ Notice that $E$ is
now a real energy, rather than the Floquet phase
it was for a map, and $T_P$ has units of time
instead of step number, hence the factor of
$\hbar$ in the equations following.}
\begin{equation} w_a \sim \hbar \lambda /T_P \,,
\end{equation}  and separated by
\begin{equation} s_a \sim \hbar /T_P \,.
\end{equation} The normalized peak height, and the
IPR are therefore given by
\begin{equation}
\label{iprhamg} {\rm IPR} \sim h_a \sim
\lambda^{-1} \,.
\end{equation} Again we see that ordinary measures
of scarring are typically unable to resolve
scarring arising from longer orbits, because the
exponent $\lambda$ for such orbits is generally
not small.

The universal state $\Psi$ is constructed by
analogy with Eq.~\ref{hiperpsi} as
\begin{eqnarray}
\Psi & = & {\cal N} \int_0^{T_P} d\tau
\int dt e^{-t^2/T_P^2 T_0^2} \nonumber \\ & \times
& e^{i (E_k + \hbar \theta/T_P)\tau/\hbar +i\theta
t/T_P -i \phi_\tau } |a_{x_\tau,\sigma_x,\sigma
f_\tau e^{\lambda t/T_P}} \rangle\,.
\end{eqnarray} Here $\tau$ is a time parameter
parametrizing the orbit $x_\tau$,
$f_\tau$ and $\phi_\tau$ are as before a
stretching factor and phase associated with short
time evolution from $x_0$ to $x_\tau$,
$E_k$ is a quantization energy, and
$\theta/\lambda$ is again an optional
off-resonance energy shift parameter. For small
$\lambda$, this can be written as
\begin{equation}
\Psi_{\rm dyn} \sim \sum_t \, e^{-t^2/T_P^2 T_0^2}
\, e^{i(E_k+\hbar \theta/T_P)t/\hbar} |a_{\rm
lin}(t) \rangle\,.
\end{equation}

The short-time autocorrelation function again has
a decay time scale
$T_P T_0 \sim T_P \lambda^{-1} \log {A/\hbar}$
(with no window $w(t)$ present), leading to a {\it
single} spectral envelope centered at energy $E_k$
($E_k+\hbar \theta/T_P$ for off-resonance states),
and of width and height given by
\begin{eqnarray} w_\Psi & \sim & \hbar \lambda /
(T_P \log {A /\hbar}) \nonumber \\ h_\Psi & \sim &
(T_P \log {A /\hbar}) / \hbar \lambda \,.
\end{eqnarray} The IPR is somewhat difficult to
talk about in this case because IPR's (like any
other measure of quantum ergodicity) can only be
measured relative to some already known energy
window which takes into account various conserved
quantities. In this case, the only plausible
window is the spectral envelope of the original
Gaussian wavepacket of width $\sigma_x$ in the
direction of the periodic orbit (see discussion
following Eq.~\ref{hamwpkt}). Then the IPR is
given by
\begin{equation} {\rm IPR}_\Psi \sim {\sqrt E T_P
\over \lambda \sigma_x}
\log{A \over \hbar} \,.
\end{equation} Notice that the combination
$\lambda / \sqrt E T_P$ is just the exponent per
unit length of the orbit. Of course, our result
depends on the width $\sigma_x$ of each Gaussian
along the orbit. The enhancement of the IPR for
$\Psi$ over the corresponding IPR for the single
Gaussian  (Eq.~\ref{iprhamg}) can also be thought
of as being given by the usual logarithmic factor
times the ratio $\sqrt E T_P / \sigma_x$ of the
length of the orbit compared to the wavepacket
size.

In any case, we again see that long orbits can be
easily resolved using this improved scarring
measure. A sufficient condition to get significant
enhancement is for the exponential stretching to
be small on the time scale of the very-short-time
window $w(t)$, i.e. during the time it takes for
the wavepacket to traverse its width while moving
along the orbit. This criterion is of course
independent of the orbit length. Even if the
criterion above is not satisfied at low energies,
one nevertheless gets strong scarring in the
semiclassical limit. For a two-dimensional
billiard system, the increase in scarring strength
(as measured by peak height of the spectrum or by
the IPR) scales with energy as
\begin{eqnarray} h_\Psi & \sim & \sqrt E \log E  \\
{\rm IPR}_\Psi & \sim & E^{1/4} \log E \,.
\end{eqnarray} (The power-law IPR scaling arises
>from the wavepacket width $\sigma_x$ scaling with
energy as $\sigma_x \sim E^{-1/4}$. This is a
natural scaling which keeps the $x-$uncertainty
and the uncertainty in the $x-$momentum $p_x$ in a fixed
ratio relative to the total size of the accessible
phase space. Without this scaling the increase in
IPR with energy would be only logarithmic.)

\subsection{Numerical tests}

\label{nump2}

We conclude this section with a numerical example
of the localization enhancement obtainable for
longer orbits using the universal test-state
approach. For this purpose, we choose a modified
version of the kicked-baker system introduced in
Section~\ref{kickbaker}. Instead of having the
kicks act on the left and right strips of the
three-strip system, we have one act now only on
the middle strip, leaving the left and right
strips to undergo ordinary baker-like dynamics,
i.e. horizontal expansion and vertical
compression. Any periodic orbits contained
entirely in these two side strips thus have
locally orthogonal stable and unstable manifolds
(of the form Eq.~\ref{eqmo}), with a stretching
exponent and action given simply in terms of the
widths $w_0$ and $w_2$ of the left and right
strips. In particular, consider the period-$2$
orbit that jumps from the left strip to the right
strip and back. Its periodic point in the left
strip is given by $q= w_0(1-w_2) / (1-w_0 w_2)$,
$p=(1-w_2) / (1-w_0 w_2)$. The other periodic
point is obtained by interchanging the $q$ and $p$
coordinates. The stretching exponent for one full
iteration of this orbit is $\lambda = |\log w_0
w_1|$, and the  corresponding phase is given by
$\phi=w_0(1-w_2)^2/(1-w_0 w_2)\hbar$.  Thus, a
desired value for the exponent and phase can be fixed by
selecting the three baker strip widths, and the
kick strength acting on the middle strip is then
used to provide ensemble averaging over the
details of the long orbits (nonlinear
recurrences). 

We select for our example widths $w_0=0.40$,
$w_2=0.42$, as before, leading to an exponent
$\lambda=|\log 0.168|$ for our chosen orbit, and
work with the matrix size $N=800$. In 
Fig.~\ref{p2specfig}, the smoothed local density
of states for a Gaussian wavepacket, obtained by
averaging over several realizations, is
represented by the double-peaked solid curve. The
theoretical prediction, given by the linearized
dynamics of Eq.~\ref{ap2}, is shown by the dashed
curve. The narrow, single-peaked solid curve
centered at one of the two possible quantization
energies is the similarly smoothed local density
of states for the universal state $\Psi$,
constructed once again with $c=0.6$ and
$\theta=0$, as in Fig.~\ref{specfig}. Again,  the
corresponding dashed curve is the theoretical
prediction based on the linearized dynamics and
agrees well with the data. Notice that in this
case the difference between Gaussian and universal
scarmometers is more dramatic than in
Fig.~\ref{specfig}, the reason being that scarring
as measured by $\Psi$ depends only on the
stretching rate along the unstable manifold per
unit time, not per iteration of the entire orbit.
While scarring strength as measured by a single
wavepacket drops off with orbit length, scarring
strength as measured by $\Psi$ is
length-independent as long the orbit period
remains small compared to the log-time. Orbits of
arbitrary length can therefore be strongly scarred
using this measure, provided a correspondingly
small value of $\hbar$ is chosen.

\section{Summary and concluding remarks}

\label{concl}

In this paper we began by reviewing the theory and
existing  measures of scarring.  We were then able 
to establish considerably refined and
arguably universal scarring measures.  The
refinement means in practice that much larger
deviations from RMT behavior are predicted
using the refined test states.  The test states
are not special from the standpoint of random 
wavefunctions, but they pick up structures 
which exist in eigenstates of dynamical
systems.  

The two major issues of generalization of scar
measures which we faced  are 1) coherence 
(or lack of it) in superpositions of localized
wavepacket  states, and 2) summing over all 
the points of periodic orbits whose period 
is greater than one iteration of a map.  The 
smooth Hamiltonian version of this is to 
coherently add up packets all along the orbit,
making phase space tubes which are related 
to the tubes of Agam and
Fishman~\cite{fishman}; see also~\cite{nlscar}.

The  universality
mentioned above stems ultimately from the use of
the linearizable domain near periodic points in
the construction of scar measures.  The scar test
states are the optimal  ones which can be
constructed with the linearized dynamics.  In
turn, we argued that the linearizable portion of 
the dynamics was a reasonable stopping point
for the definition of scar strength.  Going
beyond the linearizable dynamics is certainly
possible and semiclassically viable, but a 
problem arises in that one begins to approach
the construction of individual eigenstates, at
least in favorable cases~\cite{tomsovichell},
which is a somewhat  disturbing limit.  The reason
this is disturbing  is that such ``test'' states
for scarring pick  up (in the ideal limit) only
one state, which brands the whole eigenstate as a
scar.  Moreover, pieces of classical  manifolds
far from any given periodic orbit will have been
incorporated in the  longer time dynamics of such
a test state.  Indeed it is not at all clear that
any one periodic orbit should dominate the others
in such a state.  These new periodic orbits would
begin to  play a role in the long time dynamics
(on the  order of the Heisenberg time), so we
would not even be speaking of a scar of a given
periodic  orbit.  Given all these factors it seems
reasonable to stop at the linearizable  zone
surrounding given periodic orbits.

Various numerical tests made possible by ensemble
averaged baker map results supported the measures
established here.  The enhancements in IPR made
possible by our optimized coherent measures can be
modest  (factors of 1.5 or 2) for reasonable
$\hbar$ and  short orbits, but much more dramatic
for longer period orbits, as compared to a 
single Gaussian wavepacket measure.

Finally, we have given a theoretical basis for 
the ``off center'' scars living on the 
hyperbolic manifolds near  (but not on) a given
periodic orbit, and provided them with test states 
sensitive to their presence.

\section{Acknowledgements}

This research was supported by the National
Science Foundation under Grants   66-701-7557-2-30
and  CHE9610501.

\begin{figure}
\caption{ The correlation $C_{ab}$
(Eq.~\ref{cabdef}) between local densities of
states for two wavepackets $|a\rangle$ and
$|b\rangle$, lying on the same periodic orbit of
instability exponent $\lambda$, is plotted as a
function of width parameter $z$. The upper and
lower curves correspond to $\lambda=\log 5$ and 
$\lambda=\log 10$, respectively. The width (along
the unstable manifold) of  wavepacket $|b\rangle$
is $e^{\lambda(n+z)}$ times that of $|a\rangle$,
where $n$ is an integer. Numerical data for the
ensemble of kicked-baker systems described in
Section~\ref{kickbaker} is plotted along with the
theoretical curves from Eq.~\ref{cab}. As the
stretching factor $e^\lambda$ gets closer to
unity, so does the correlation $C_{ab}$.  }
\label{cabfig}
\end{figure}

\begin{figure}
\caption{ The minimum correlation in the local
densities of states $C_{ab}$ for two wavepackets
on an orbit of exponent $\lambda$ is plotted
versus the scarring IPR enhancement factor
$g(\lambda)$ for such an orbit. The theoretical
curve is obtained from
Eqs.~\ref{gabdef},~\ref{cab}, while the data again
comes from the ensemble of
Section~\ref{kickbaker}. The minimum
inter-wavepacket correlation gets very close to
unity for significant scarring enhancement factors
$g(\lambda)$. }
\label{cabfig2}
\end{figure}

\begin{figure}
\caption{ Smoothed local densities of states are
plotted for the universal hyperbolic test state
$|\Psi\rangle$ (higher peak) and a Gaussian
wavepacket $|a\rangle$ (lower peak), on a periodic
orbit with exponent
$\lambda = |\log 0.18|$. The system size is
$N=800$. Cut-off constant $c=0.6$ (see
Eq.~\ref{t0def})  is used to construct the state
$|\Psi\rangle$. The theoretical curves (dashed)
are obtained by fourier transforming the
linearized autocorrelation functions of
Eqs.~\ref{apsi},~\ref{shortcorr}, respectively,
while the data (solid curves) is obtained by
ensemble averaging. }
\label{specfig}
\end{figure}

\begin{figure}
\caption{ In this figure, the inverse
participation ratio (IPR) for hyperbolic test state
$|\Psi\rangle$ is plotted versus the log-time
cutoff $T_0$ (see Eqs.~\ref{cohdef},~\ref{t0def}),
for various values of system size
$N$. From bottom to top, the five curves
correspond to $N=50$, $100$, $200$,
$400$, and $800$. For each $N$, $26$ points are
plotted, for $c=(1.1)^j$,
$j=-20 \ldots +5$. The orbit has exponent $\lambda
= |\log 0.18|$, as in the previous figure. The
upper dashed curve is the $N \to \infty$
theoretical prediction (Eq.~\ref{iprpsi}, $F=2$),
which converges to the asymptotic prediction of
Eq.~\ref{iprpsi2} (lower dashed line) for large
$T_0$. The linearized theory is expected to start
breaking down for $c \ge 1$ (rightmost six points
on each data curve). }
\label{iprfig}
\end{figure}

\begin{figure}
\caption{ This figure is a Husimi plot in phase
space of the universal hyperbolic state
$|\Psi\rangle$ for log-time cutoff $T \to \infty$
(i.e. the region plotted is well inside the
linearizable region in classical phase space). The
fixed point it at the center of the plot, and the
horizontal and vertical axes correspond to the
unstable and stable manifolds, respectively. The
total area of the plot is $12 \sqrt\hbar \times 12
\sqrt\hbar$. }
\label{husfig}
\end{figure}

\begin{figure}
\caption{ Smoothed local densities of states are
plotted for the off-resonance universal hyperbolic
test state $|\Psi\rangle$, for off-resonance angle
$\theta$ ranging from $0$ (tallest peak) in steps
of $\pi/10$ (to the left), through $-2\pi$, on a
periodic orbit with exponent
$\lambda = |\log 0.18|$. As in Fig.~\ref{specfig},
the system size is $N=800$, and the cut-off
constant $c$ is set to $0.6$ (see Eq.~\ref{t0def}).
The data is obtained by ensemble averaging. }
\label{specfigth}
\end{figure}

\begin{figure}
\caption{ Theoretically computed IPR values for
the universal hyperbolic test states
$\Psi$ are plotted versus the log-time $T_0$  for
several values of off-resonance angle $\theta$.
From top to bottom at $T_0=1$, the solid curves
represent $\theta=0 \ldots \pi$, in steps of
$\pi/5$.  The dashed line is the limiting value
for all of these at large $T_0$
(Eq.~\ref{iprpsi2}). The single-Gaussian IPR (the
$T_0 \to 0$ limit) is $3.66$ for this value of
$\lambda$. }
\label{iprthfig}
\end{figure}

\begin{figure}
\caption{ The IPR is plotted here as a function of
log-time $T_0$, as in Fig.~\ref{iprfig}, but for
the off-resonance test state $\Psi(\theta=\pi)$ of
Eq.~\ref{cohdefth}. Again, five data curves
corresponding (bottom to top) to
$N=50$, $N=100$, $N=200$, $N=400$, and $N=800$ are
shown in the Figure, with cutoff parameter $c$
varying from $(1.1)^{-20}$ to $(1.1)^{+5}$ from
left to right within each curve. The stability
exponent is $\lambda=|\log 0.18|$, as before. The
upper dashed curve is the $N \to \infty$
theoretical prediction obtained from
Eqs.~\ref{iprpsi},~\ref{apsith}. The lower dashed
line is the asymptotic form of Eq.~\ref{iprpsi2},
to which the upper curve converges in the $T_0 \to
\infty$ limit. The linearized theory is again
expected to break down for $c \ge 1$ (the
rightmost six points on each data curve). }
\label{iprpifig}
\end{figure}

\begin{figure}
\caption{ Here the IPR is plotted versus
off-resonance angle $\theta$, for
$\lambda=|\log 0.18|$, $N=200$ and cutoff time
parameter $c=0.6$. The solid curve with errorbars
represents ensemble-averaged data, while the
dashed curve is the theory
(Eqs.~\ref{iprpsi},~\ref{apsith}). At small
angles, the scarring strength is significantly
above the single wavepacket IPR value of $3.5$
(theoretical prediction:
$3.66$), while for larger angles we approach the
ergodic RMT value of $2$. }
\label{iprvsth}
\end{figure}

\begin{figure}
\caption{ Analogous to Fig.~\ref{husfig}, this
figure presents Husimi plots of the universal test
state $\Psi$ for off-resonance parameter values
(a) $\theta/\lambda=0.8$ and (b)
$\theta/\lambda=2.5$. As before, the linearizable
region is taken to be much larger than the
displayed area of size  $12 \sqrt\hbar \times 12
\sqrt\hbar$, centered on the periodic orbit. }
\label{husfigth}
\end{figure}

\begin{figure}
\caption{ Smoothed local densities of states are
shown here for a Gaussian wavepacket placed on a
period-$2$ orbit (double-peaked solid curve), and
the universal test state $\Psi$ constructed on the
same orbit (tall single peak). The dashed curves
represent theoretical predictions based on the
linearized dynamics near the periodic  orbit in
question. The system is a kicked baker's map with
kick potential acting on the middle strip (see
Section~\ref{nump2}), and the periodic orbit has a
total exponent $\lambda=|\log 0.168|$ over the
two-step period. One of two possible on-resonance
energies has been chosen for the test state
$\Psi$, which is again constructed using cutoff
constant
$c=0.6$ (as in the analogous calculation in
Fig.~\ref{specfig} for a period-one orbit). The
enhancement here is more dramatic due to the fact
that universal scarring strength depends only on
the exponent per unit time along the orbit, not on
the orbit length itself. }
\label{p2specfig}
\end{figure}


\begin{references}

\bibitem{bohigas} O. Bohigas, M.-J. Giannoni, and
C. Schmit, {\it J. Physique Lett.} {\bf 45},
L-1015 (1984).

\bibitem{hel84}  E. J. Heller,  {\it Phys. Rev.
Lett. } {\bf 53}, 1515-18 (1984).

\bibitem{thesis} S. W. McDonald, Ph.D. thesis,
Lawrence Berkeley Laboratory, LBL14837 (1983).

\bibitem{mk}   S. W. McDonald and A. N. Kaufman,
{\it Phys. Rev. Lett. } {\bf 42}, 1189  (1979).

\bibitem{berry}   M. V. Berry, {\it Ann. NY Acad.
Sci.}  {\bf 357}, 183 (1983).

\bibitem{bogo} E. B. Bogomolny, {\it Physica} {\bf
D 31}, 169 (1988).

\bibitem{berryscar} M. V. Berry,  Les Houches
Lecture Notes, Summer School on Chaos and Quantum
Physics,  M.-J. Giannoni, A. Voros, and J.
Zinn-Justin, eds., Elsevier Science Publishers
B.V. (1991); M. V. Berry, {\it Proc. Roy. Soc.} 
{\bf A 243}, 219 (1989).

\bibitem{scarexp} A. D. Peters, C. Jaffe, and J.
Delos, {\it Phys. Rev. } {\bf A 56}, 311 (1997). 

\bibitem{diode}
T. M. Fromhold, P. B. Wilkinson, F. W. Sheard, L. Eaves, J. Miao, and
G. Edwards, {\it Phys. Rev. Lett.} {\bf 75}, 1142
(1995);   P. B. Wilkinson, T. M.  Fromhold, L. 
Eaves, F. W.  Sheard, N. Miura, and T.
       Takamasu, {\it Nature}, {\bf 380}, 608
(1996).

\bibitem{kapdecay} L. Kaplan, ``Scar and antiscar quantum effects
in open chaotic systems," preprint.

\bibitem{helocon}   P. O'Connor, J. N. Gehlen and
E. J. Heller,  {\it Phys. Rev. Lett. } {\bf 58},
1296 (1987).

\bibitem{nlscar}  L. Kaplan and E. J. Heller, {\it
Ann. Phys. (N. Y.)} {\bf 264}, 171 (1998).

\bibitem{sscar} L. Kaplan, {\it Phys. Rev. Lett.}
{\bf 80}, 2582 (1998).

\bibitem{kaplan2} L. Kaplan, ``Recent developments in the theory
of scarring," to appear in {\it Nonlinearity}.

\bibitem{schnirl} A. I. Schnirelman, {\it Usp.
Mat. Nauk.} {\bf 29}, 181 (1974).

\bibitem{zel} S. Zelditch, {\it Duke Math. J.}
{\bf 55}, 919 (1987).

\bibitem{col} Y. Colin de Verdiere, {\it Commun.
Math. Phys.}  {\bf 102}, 497 (1985).

\bibitem{tomsovichell} S. Tomsovic and E. J.
Heller,  {\it Phys. Rev. Lett. } {\bf 70}, 1405
(1993).

\bibitem{ltsc} L. Kaplan, 
{\it Phys. Rev.} {\bf E 58}, 2983 (1998).

\bibitem{ott} T. M. Antonsen, Jr., E. Ott, Q.
Chen, and R. N. Oerter, {\it Phys. Rev.} {\bf E
51}, 111 (1995).

\bibitem{caustic} M. A. Sepulveda, S. Tomsovic, and
E. J. Heller, {\it Phys. Rev. Lett.} {\bf 69}, 402
(1992).

\bibitem{fishman} O. Agam and S. Fishman, {\it
Phys. Rev. Lett.} {\bf 73}, 806 (1994); O. Agam
and S. Fishman, {\it J. Phys.} {\bf A 26}, 2113
(1993); O. Agam and N. Brenner, {\it J. Phys.}
{\bf A 28}, 1345 (1995); S. Fishman, B. Georgeot,
and R. E. Prange, {\it J. Phys.} {\bf A 29}, 919
(1996).

\bibitem{bambi} B. Li   and B. Hu   {\it J. Phys.}
{\bf A 31}, 483 (1998).

\bibitem{borondo} G. G. de Polavieja, F. Borondo,
and R. M. Benito, {\it Phys. Rev. Lett.} {\bf 73},
1613 (1994).

\bibitem{smilansky} D. Klakow and U. Smilansky,
{\it J. Phys.} {\bf A 29}, 3213 (1996).

\bibitem{dealmeida} A. M. Ozorio de Almeida, {\it
Phys. Rep.} {\bf 295}, 265 (1998).

\bibitem{tomsovic} S. Tomsovic, {\it Phys. Rev.
Lett.} {\bf 77}, 4158 (1996).

\bibitem{arranz} F. J. Arranz, F. Borondo, and R.
M. Benito, {\it Phys. Rev. Lett.} {\bf 80}, 944
(1998).

\bibitem{voros} S. Nonnenmacher and A. Voros, {\it
J. Phys.} {\bf A 30}, 295 (1997).

\bibitem{saraceno} F. P. Simonotti, E. Vergini,
and M. Saraceno, {\it Phys. Rev.} {\bf E 56}, 3859
(1997).

\bibitem{baker} N. L. Balazs and  A. Voros, {\it
Europhys. Lett.}
 {\bf 4}, 1089 (1987);  N. L. Balazs and  A. Voros,
{\it Ann. Phys. (N. Y.)} {\bf 190}, 1 (1989); M.
Saraceno, {\it Ann. Phys. (N. Y.)} {\bf 199}, 37
(1990);
 M. Saraceno and A. Voros, {\it Physica} {\bf D
79}, 206 (1994).

\bibitem {wqe}  L. Kaplan and E. J. Heller,
{\it Physica} {\bf D 121}, 1 (1998).

\bibitem{kickmap} G. Casati, B. V. Chirikov, F. M.
Izrailev, and J. Ford, in {\it Stochastic Behavior
in Classical and Quantum Hamiltonian Systems}, ed.
by G. Casati and J. Ford, Springer, New York
(1979); M. V. Berry, N. L. Balazs, M. Tabor, and
A. Voros, {\it Ann. Phys. (N.Y.)} {\bf 122}, 122
(1979).

\end{references}
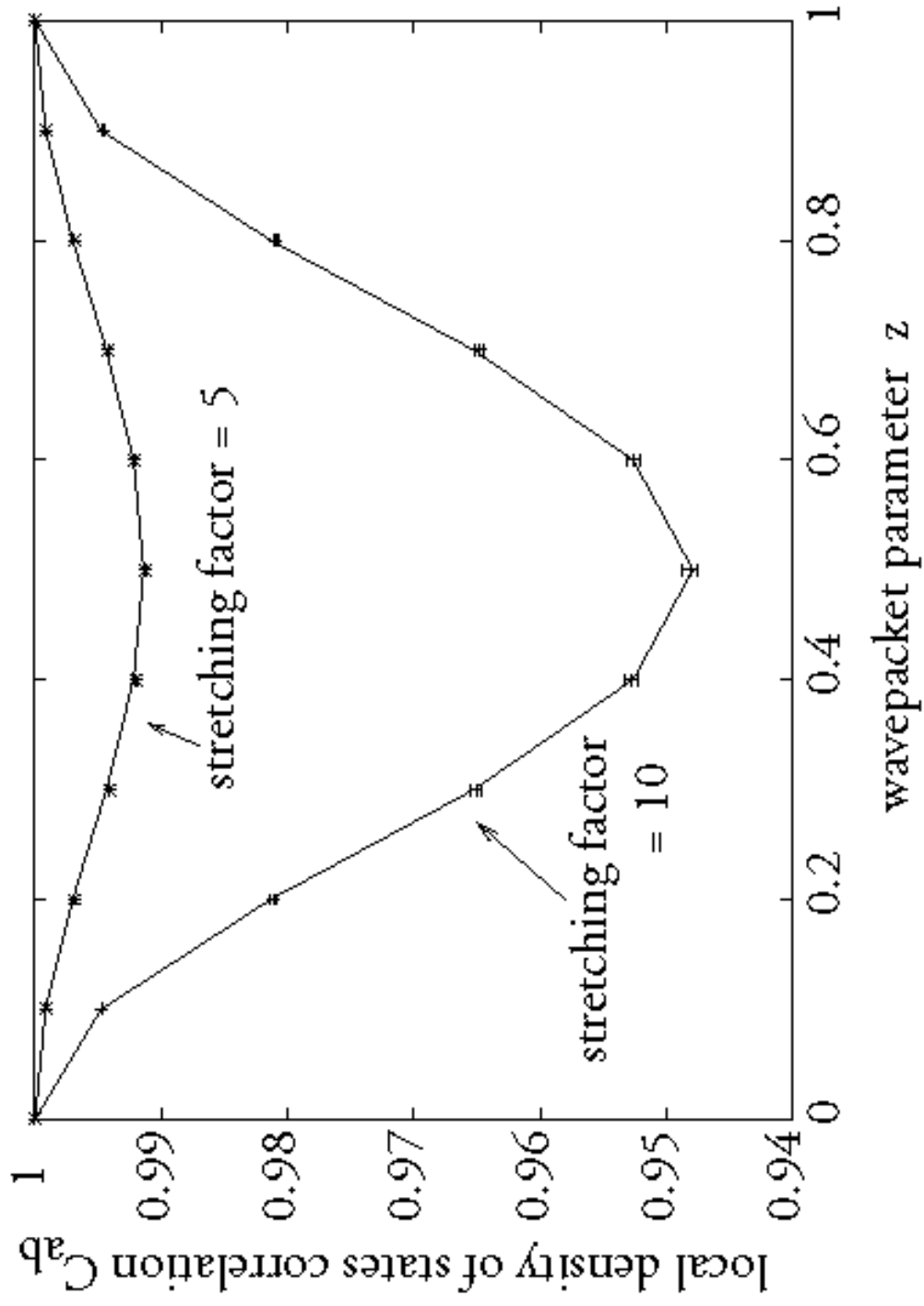
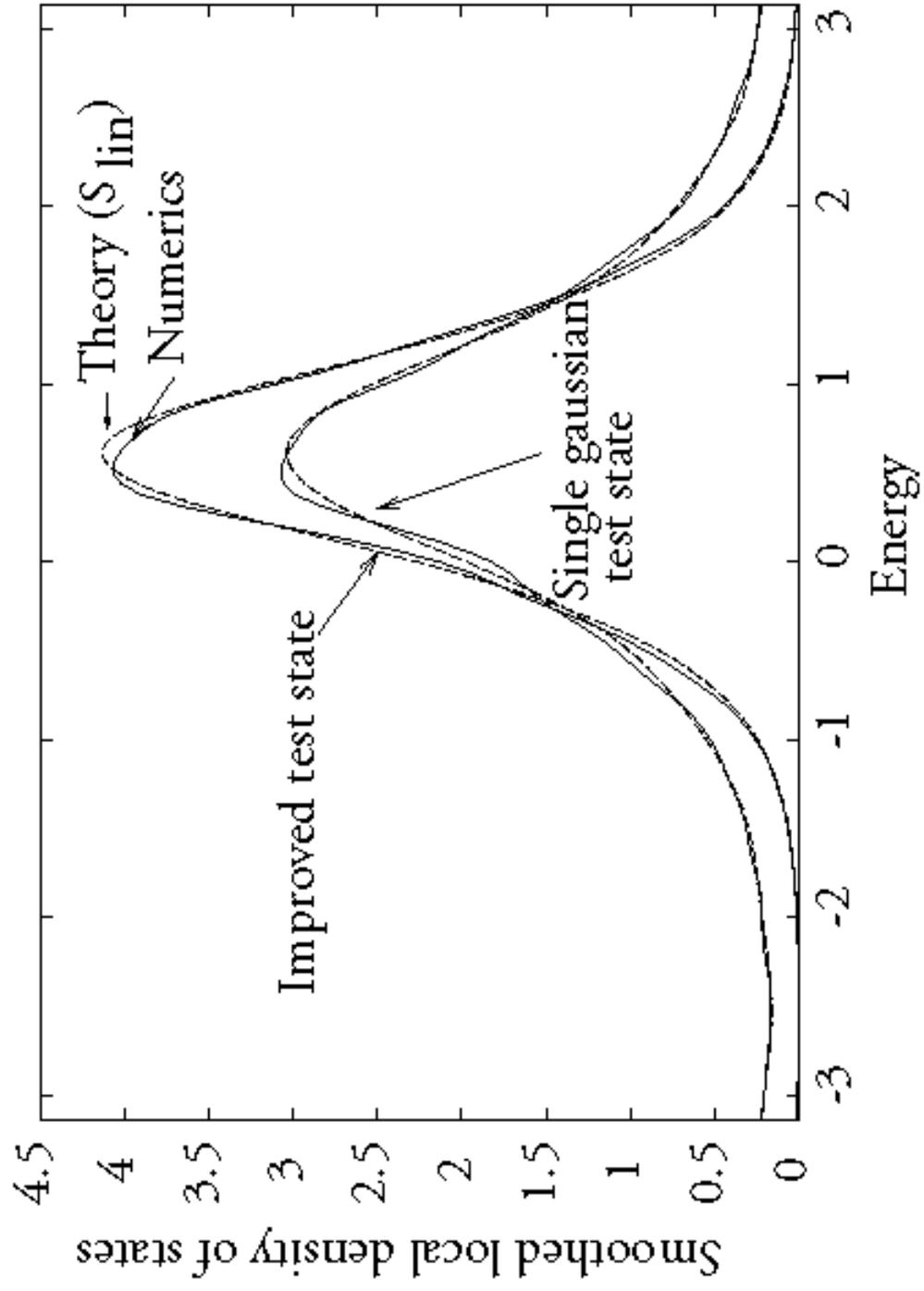
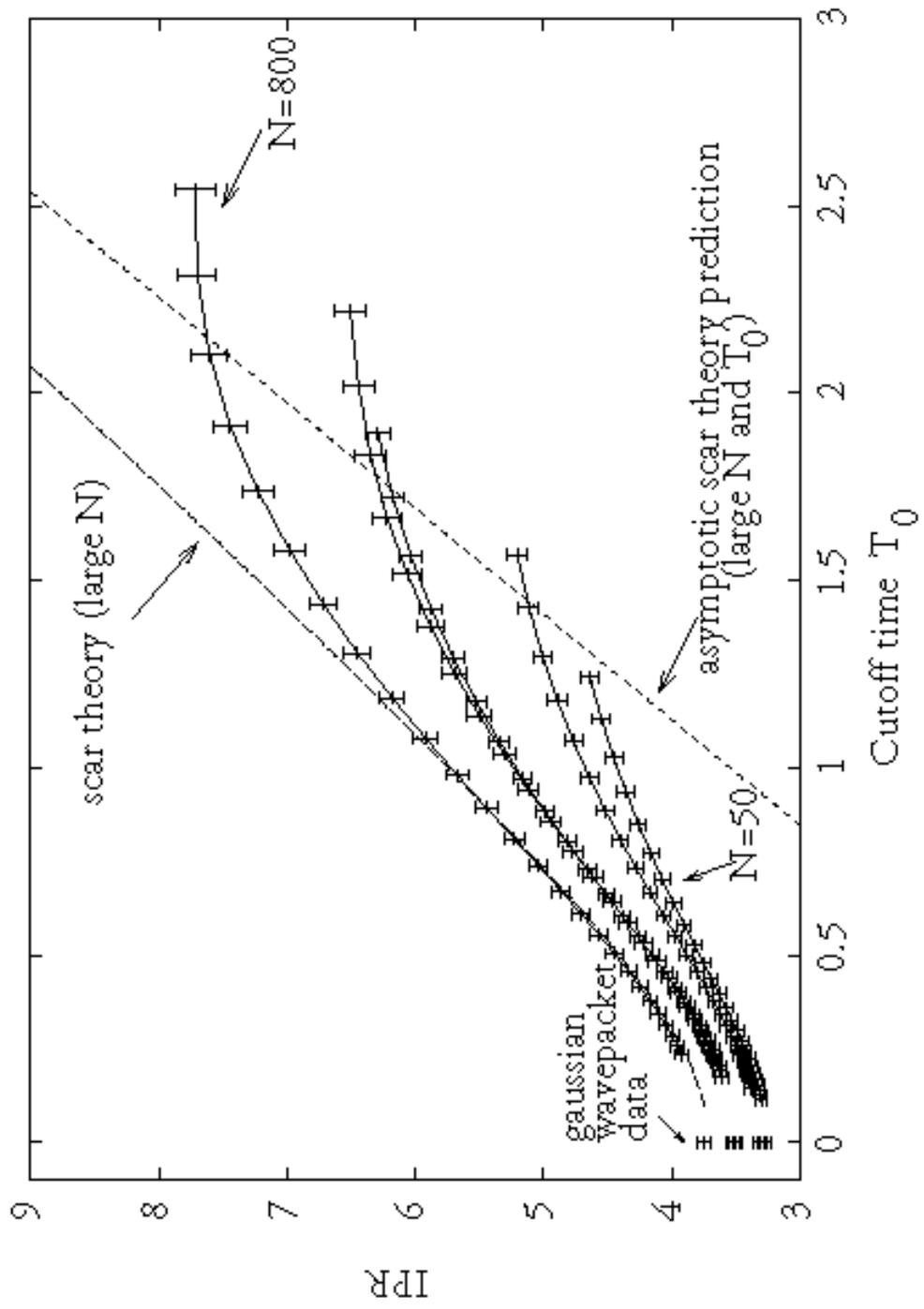
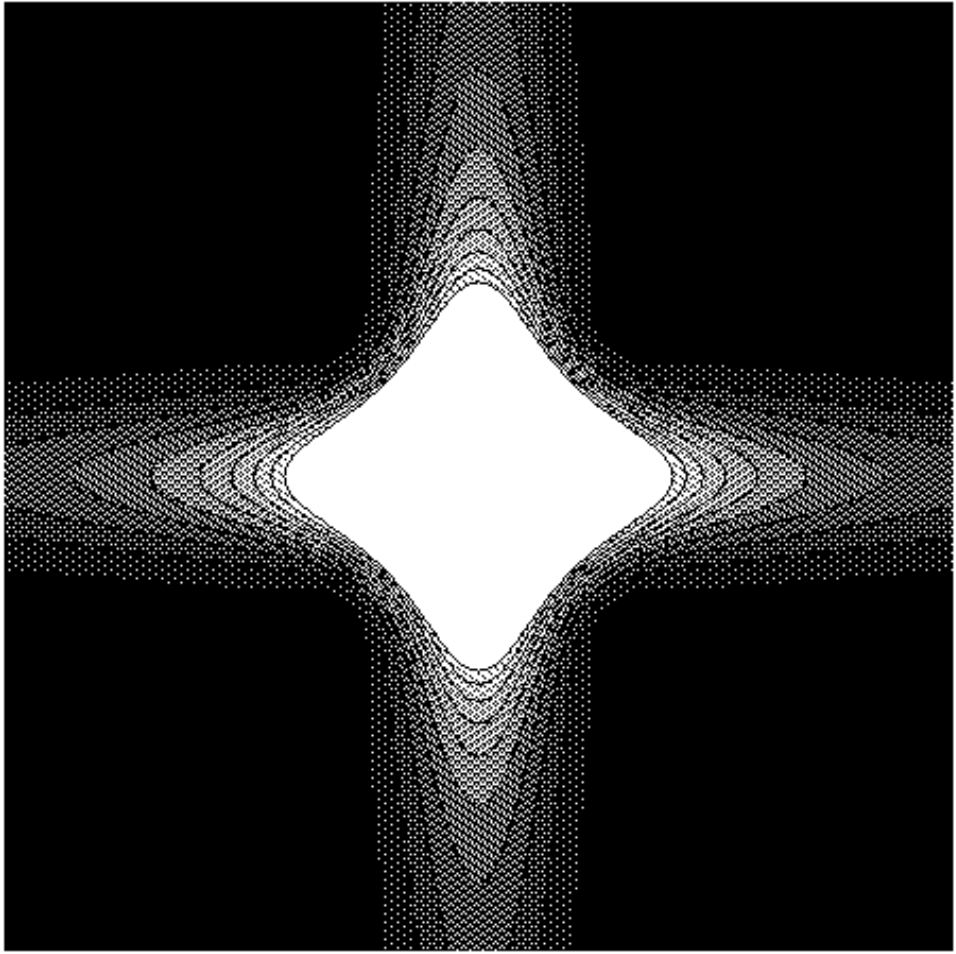
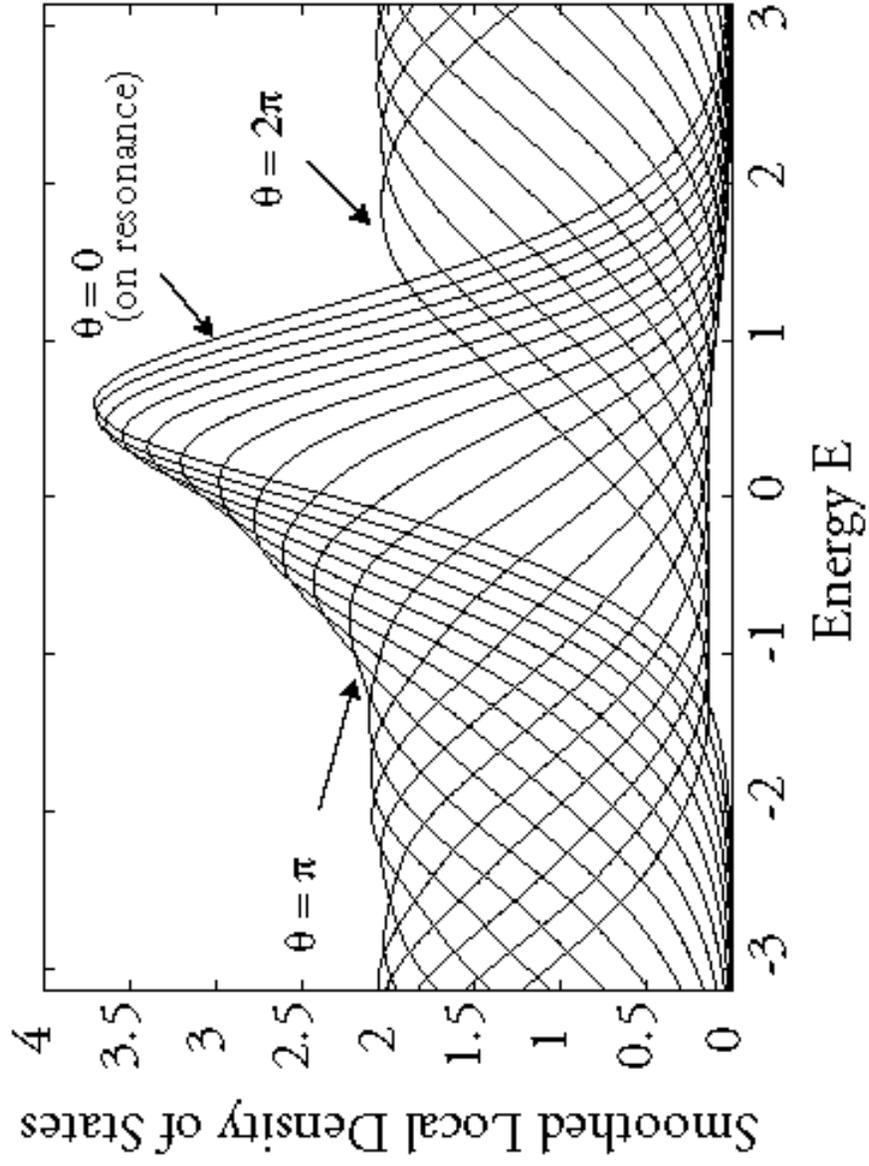
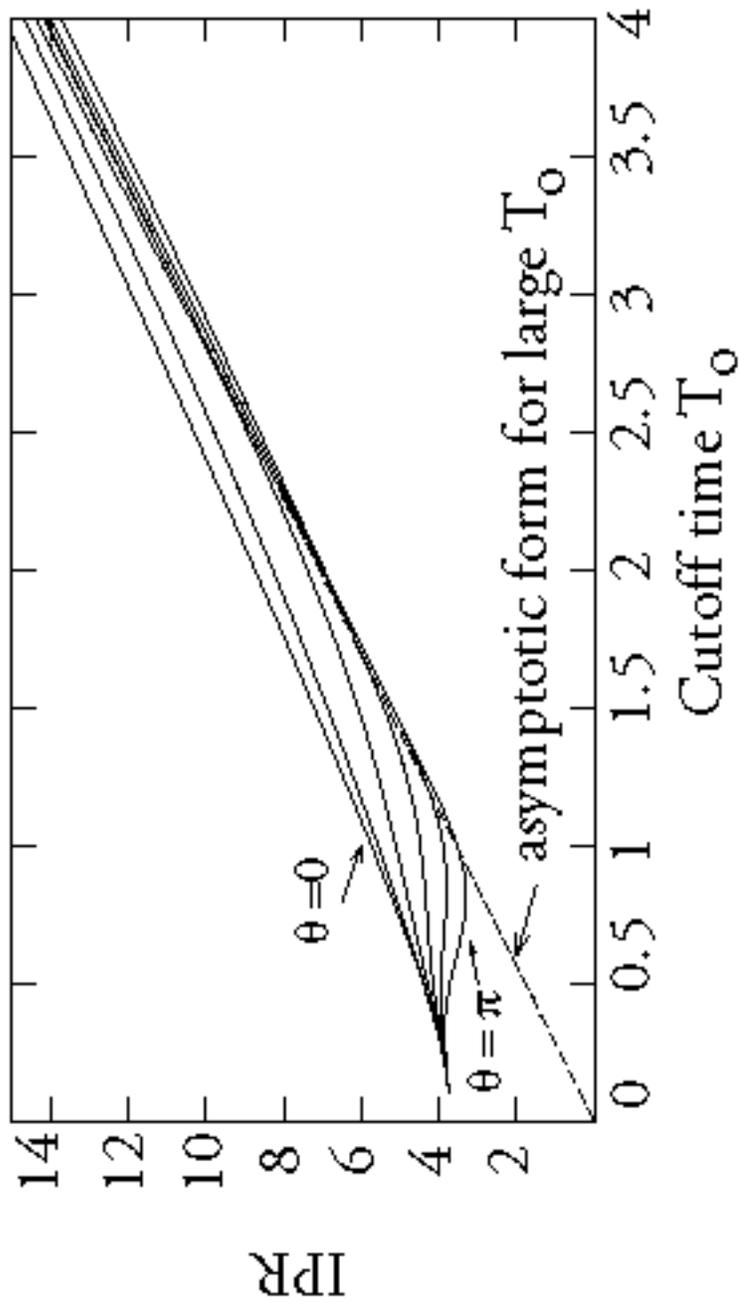
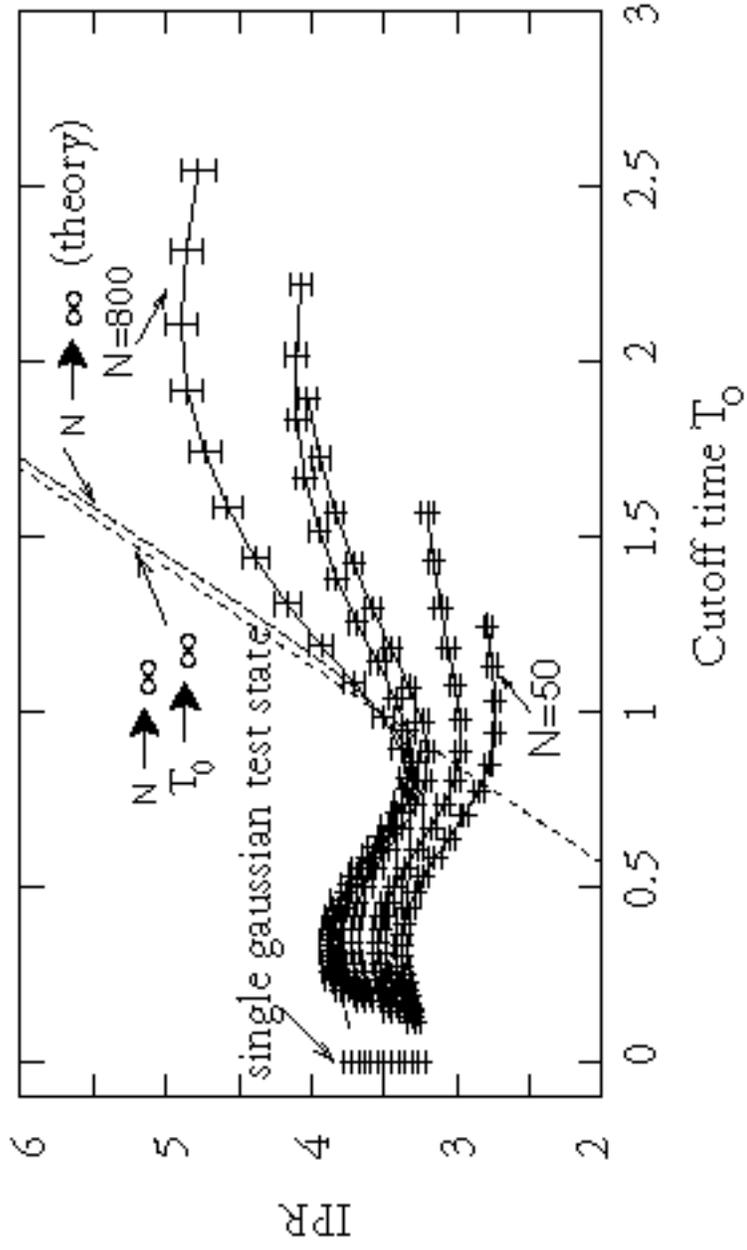
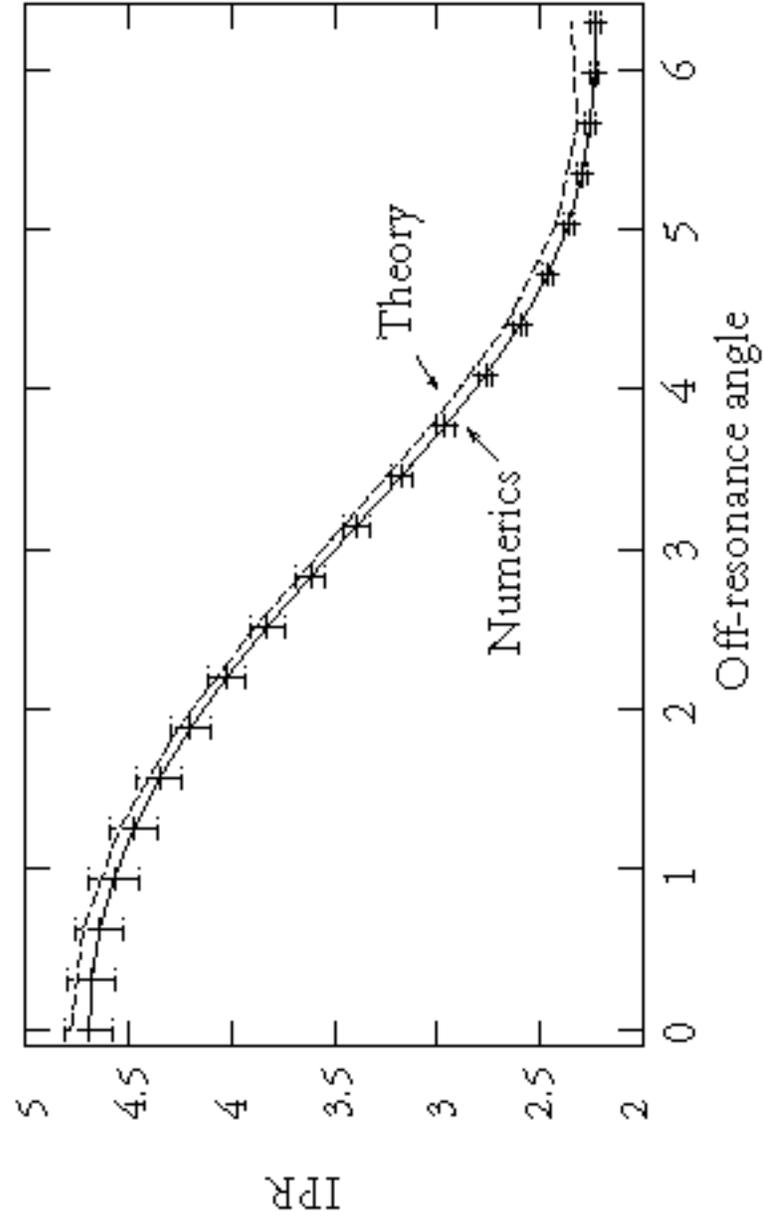
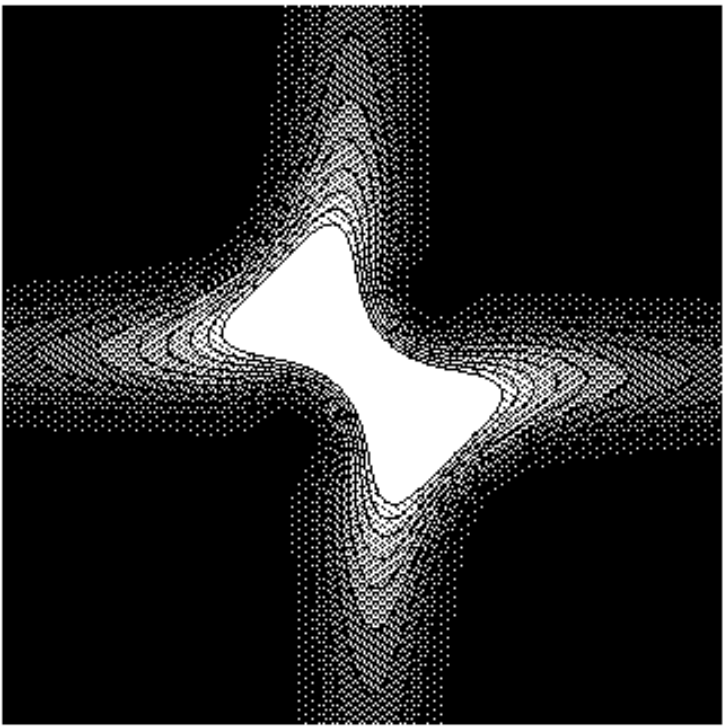
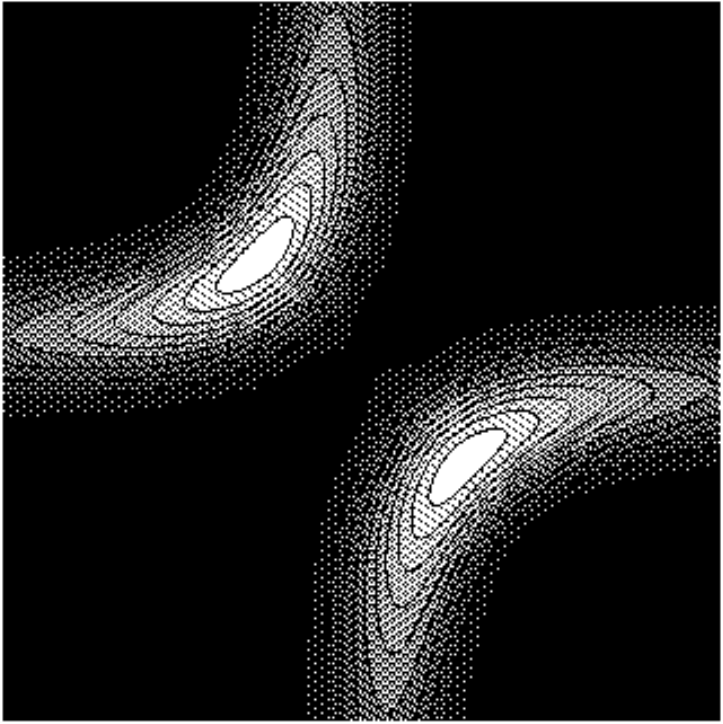
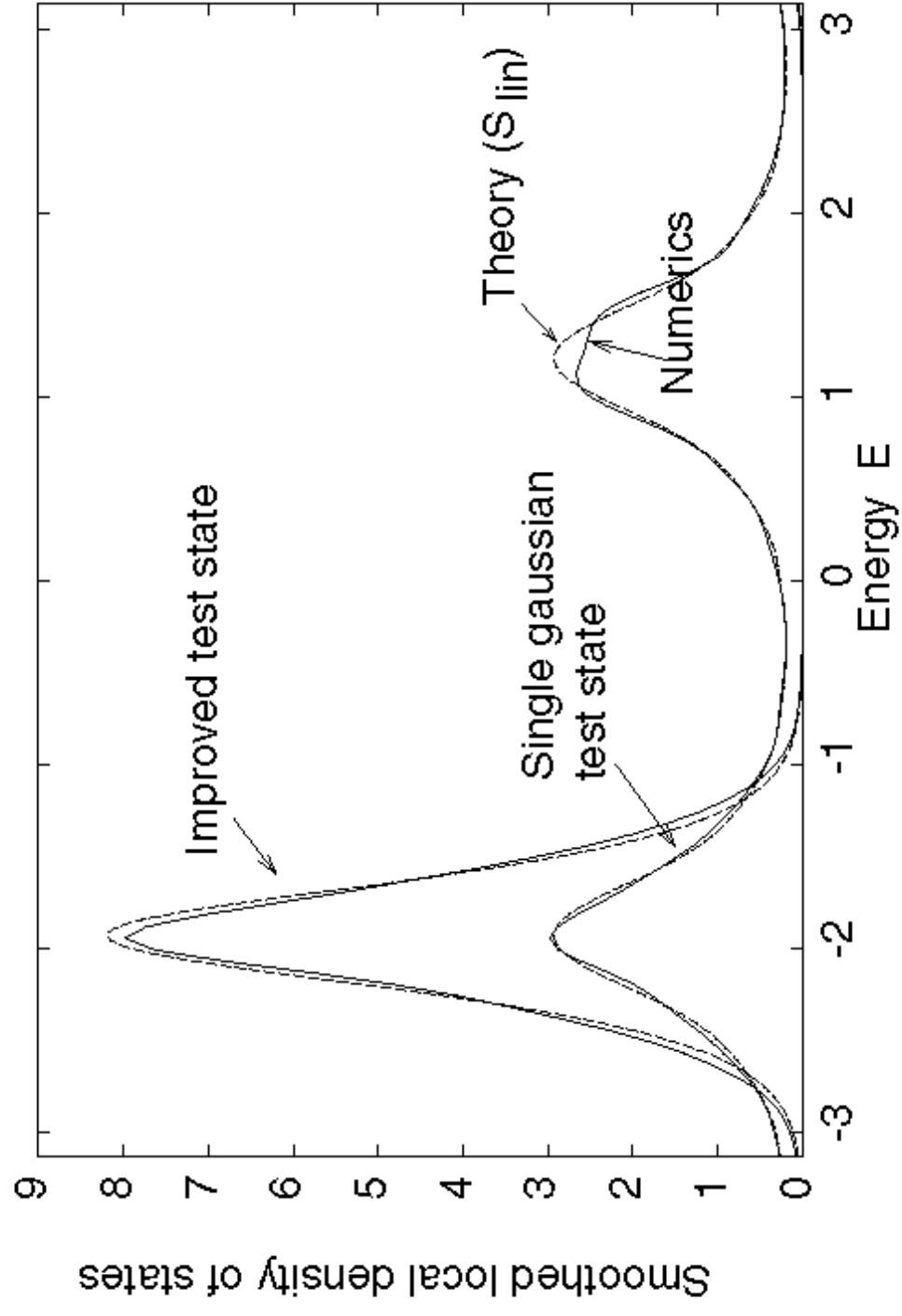
\end{document}